\documentclass[aps,reprint,prl,twocolumn,superscriptaddress]{revtex4-1}
\usepackage{graphicx}
\usepackage[colorlinks,urlcolor=blue,citecolor=blue,linkcolor=blue]{hyperref}
\usepackage{color}
\usepackage{float}
\usepackage{amsmath}
\usepackage{amssymb}
\usepackage{CJK}

\begin{document}
\hyphenation{Fesh-bach}

\begin{CJK*}{UTF8}{gbsn} 
\title{Resonantly Interacting Fermi-Fermi Mixture of $^{161}$Dy and $^{40}$K}

\author{C. Ravensbergen}
\affiliation{Institut f{\"u}r Experimentalphysik, Universit{\"a}t Innsbruck, 6020 Innsbruck, Austria}
\affiliation{Institut f{\"u}r Quantenoptik und Quanteninformation (IQOQI), {\"O}sterreichische Akademie der Wissenschaften, 6020 Innsbruck, Austria}

\author{E. Soave}
\affiliation{Institut f{\"u}r Experimentalphysik, Universit{\"a}t Innsbruck, 6020 Innsbruck, Austria}

\author{V. Corre}
\affiliation{Institut f{\"u}r Experimentalphysik, Universit{\"a}t Innsbruck, 6020 Innsbruck, Austria}
\affiliation{Institut f{\"u}r Quantenoptik und Quanteninformation (IQOQI), {\"O}sterreichische Akademie der Wissenschaften, 6020 Innsbruck, Austria}

\author{M. Kreyer}
\affiliation{Institut f{\"u}r Experimentalphysik, Universit{\"a}t Innsbruck, 6020 Innsbruck, Austria}

\author{B. Huang (黄博)} 
\affiliation{Institut f{\"u}r Experimentalphysik, Universit{\"a}t Innsbruck, 6020 Innsbruck, Austria}
\affiliation{Institut f{\"u}r Quantenoptik und Quanteninformation (IQOQI), {\"O}sterreichische Akademie der Wissenschaften, 6020 Innsbruck, Austria}

\author{E. Kirilov}
\affiliation{Institut f{\"u}r Experimentalphysik, Universit{\"a}t Innsbruck, 6020 Innsbruck, Austria}

\author{R. Grimm}
\affiliation{Institut f{\"u}r Experimentalphysik, Universit{\"a}t Innsbruck, 6020 Innsbruck, Austria}
\affiliation{Institut f{\"u}r Quantenoptik und Quanteninformation (IQOQI), {\"O}sterreichische Akademie der Wissenschaften, 6020 Innsbruck, Austria}

\date{\today}

\begin{abstract}
We report on the realization of a Fermi-Fermi mixture of ultracold atoms that combines mass imbalance, tunability, and collisional stability. In an optically trapped sample of $^{161}$Dy and $^{40}$K, we identify a broad Feshbach resonance centered at a magnetic field of {217\,G}. Hydrodynamic expansion profiles in the resonant interaction regime reveal a bimodal behavior resulting from mass imbalance. Lifetime studies on resonance show a suppression of inelastic few-body processes by orders of magnitude, which we interpret as a consequence of the fermionic nature of our system. The resonant mixture opens up intriguing perspectives for studies on novel states of strongly correlated fermions with mass imbalance.
\end{abstract}


\maketitle
\end{CJK*}

Ultracold Fermi gases with {resonant} interactions have attracted a great deal of attention  as precisely controllable model systems for quantum many-body physics \cite{Inguscio2006ufg,Pitaevskii2016book,Zwerger2012tbb,Strinati2018tbb}. The interest spans across many different fields, from primordial matter, neutron stars and atomic nuclei to condensed-matter systems, and in particular concerning superfluids and superconductors \cite{Bennemann2014both}. Corresponding experiments in ultracold Fermi gases require strong $s$-wave interactions, which can be realized based on Feshbach resonances \cite{Chin2010fri} in two-component systems. The vast majority of experiments in this field relies on spin mixtures of fermionic atomic species, which naturally imposes equal masses. Beyond this well-established situation, theoretical work has predicted fermionic systems with mass imbalance to favor exotic interaction regimes \cite{Gubbels2013ifg}. Mass-imbalanced systems hold particular promise \cite{Gubbels2009lpi,Wang2017eeo} in view of superfluid states with unconventional pairing mechanisms, most notably the elusive Fulde-Ferrell-Larkin-Ovchinnikov (FFLO) state \cite{Fulde1964sia,Larkin1964iso,radzihovsky2010ifr}.

A key factor for experiments on resonantly interacting Fermi gases is the collisional stability that arises from a suppression of inelastic loss processes at large scattering lengths. This effect is a result of Pauli exclusion in few-body processes at ultralow energies \cite{Petrov2004wbd,Petrov2005spo}. To act efficiently in an experiment, the suppression requires a broad $s$-wave Feshbach resonance with a sufficiently large universal range \cite{Petrov2005spo,Levinsen2011ada}. 
For the mass-balanced case, suitable resonances exist in spin mixtures of $^6$Li or $^{40}$K, and such systems are used in many laboratories worldwide. In a mass-imbalanced fermion system, the same suppression effect can be expected \cite{Petrov2005dmi}. However,  the only $s$-wave tunable Fermi-Fermi system realized so far is the mixture of  $^6$Li and $^{40}$K \cite{Trenkwalder2011heo,Jag2014ooa}, for which the Feshbach resonances \cite{Wille2008eau,Tiecke2010bfr,Naik2011fri} are too narrow to enable strong loss suppression \cite{Jag2016lof}. 

The advent of submerged-shell lanthanide atoms  in the field of ultracold quantum gases \cite{Lu2011sdb,Aikawa2012bec,Lu2012qdd,Aikawa2014rfd} has considerably enhanced the experimental possibilities. While most of the current work focuses on interactions that result from the large magnetic dipole moment or the complex optical transition structure, the availability of additional fermionic atoms is of great interest in view of novel ultracold mixtures and strongly interacting systems \cite{Trautmann2018dqm,Baier2018roa}. We have recently introduced the mixture of $^{161}$Dy and $^{40}$K \cite{Ravensbergen2018ado,Ravensbergen2018poa} as a candidate for realizing a collisionally stable, strongly interacting Fermi-Fermi mixture. Many narrow Feshbach resonances can be expected for such a system as a result of anisotropic interatomic interactions \cite{Petrov2012aif,Gonzalezmartinez2015mtf}. However, the key question in view of future experiments has remained, whether suitable broad resonances would exist. 

\begin{figure}
\includegraphics[width=1\columnwidth]{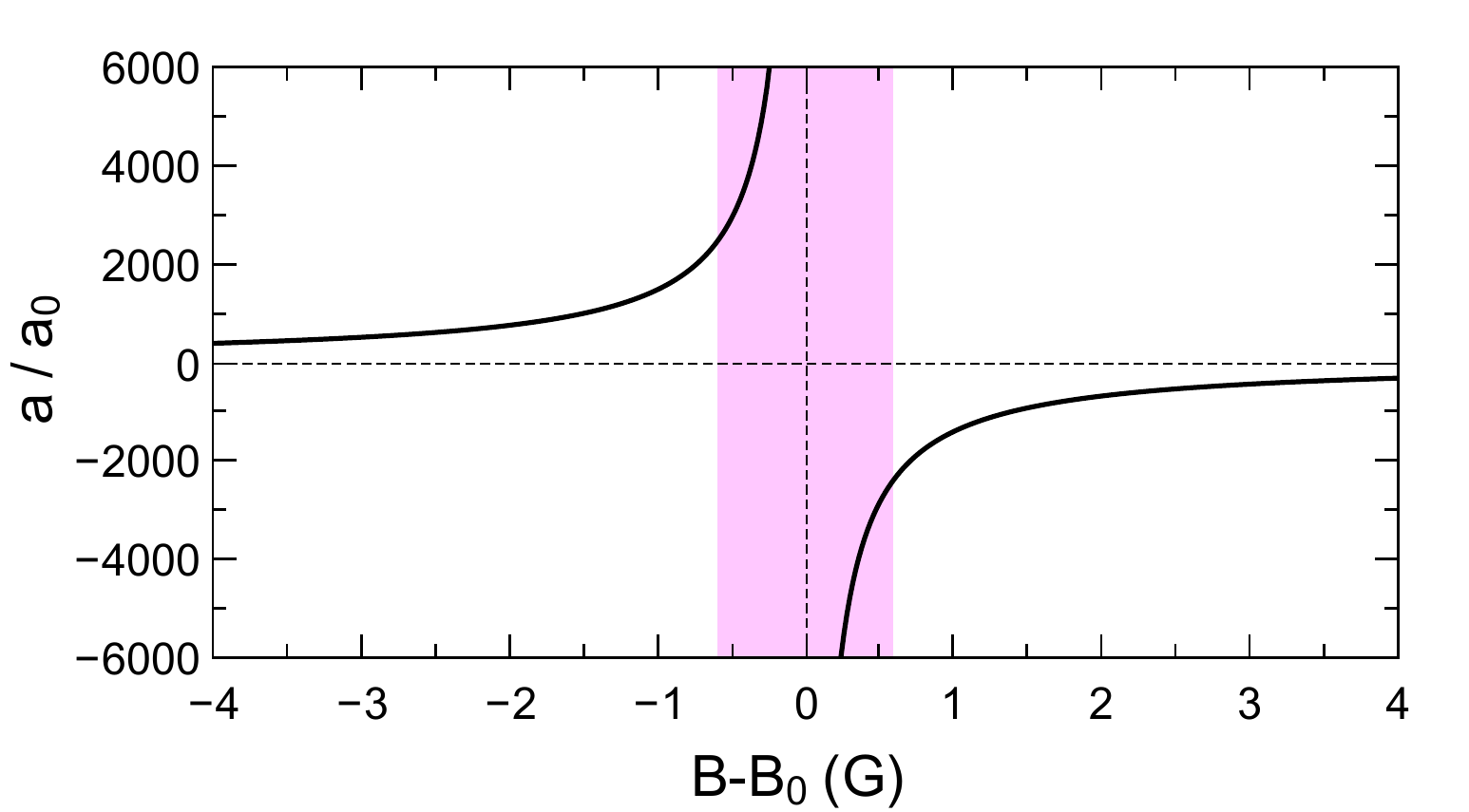}
\caption{Interspecies scattering length $a$ for $^{161}$Dy-$^{40}$K near the broad Feshbach resonance centered at $B_0 \approx 217$\,G. The shaded region indicates the regime where $a$ exceeds all other relevant length scales (see text). 
}
\label{Fig1}
\end{figure}  

In this Letter, we report on a broad Feshbach resonance in the $^{161}$Dy-$^{40}$K mixture with its center found near 217\,G. We have identified this resonance (see Fig.~\ref{Fig1}) as the strongest one in a scenario of three overlapping resonances \cite{SM}, with the other two at 200\,G and 253\,G. Some weak (only few mG wide) interspecies resonances do also exist in the relevant region, but they can be ignored for understanding the general structure of the broad scenario. We have characterized the three resonances by interspecies thermalization measurements, as reported in detail in \cite{SM}. Close to the center of the strongest resonance,  the tunability of the interspecies $s$-wave scattering length can be well approximated by
\begin{equation}
a = -\frac{A}{B-B_0} \, a_0 \, ,
\label{eq:a}
\end{equation}
where $a_0$ is Bohr's radius. Our best knowledge of the pole position and the strength parameter is $B_0 = 217.27(15)\,$G and $A = 1450(230)$\,G \cite{SM}. 

The starting point of our experiments is a degenerate mixture of  $^{161}$Dy and $^{40}$K, prepared in crossed-beam optical dipole trap  according to the procedures described in our earlier work \cite{Ravensbergen2018poa}. Evaporative cooling is performed at a low magnetic field of 225\,mG. Both species are in their lowest hyperfine and Zeeman substates, which excludes two-body losses. The transfer of the system into the high-field region above 200\,G is challenging, because many Dy intraspecies \cite{Baumann2014ool,Burdick2016lls} and Dy-K interspecies resonances have to be crossed in a fast ramp of the magnetic field. To minimize unwanted losses, heating, and excitations of the trapped cloud we proceed in two steps. 
Within a few ms, we ramp up the magnetic bias field to 219.6\,G \cite{settle}, where the system is given a time of a few 10\,ms to settle and establish thermal equilibrium. We then apply a very fast (2-ms) small-amplitude ramp to the target field, where the experiments are carried out. Throughout the whole sequence after evaporation, a magnetic levitation field is applied to compensate for the relative gravitational sag of both species \cite{Ravensbergen2018poa}.
In this way, we reach typical conditions of $N_{\rm Dy} = 20\,000$ and $N_{\rm K} = 8\,000$  atoms at a temperature of $T = 500$\,nK \cite{thermometry} in a slightly elongated trap (aspect ratio $\sim$2) with mean oscillation frequencies of $\bar{\omega}_{\rm Dy}/2\pi = 120$\,Hz and $\bar{\omega}_{\rm K}/2\pi = 430$\,Hz \cite{odt} and depths corresponding to $3.5\,\mu$K and $10\,\mu$K, respectively. With Fermi temperatures  of $T_F^{\rm Dy} = 290\,$nK and  $T_F^{\rm K} = 750\,$nK, our experimental conditions are near-degenerate of ($T/T_F^{\rm Dy} = 1.7$ and $T/T_F^{\rm K} = 0.65$). 

Interaction regimes near resonance can be discussed by comparing the scattering length with other relevant length scales. To characterize the interaction strength on resonance, where scattering is limited by unitarity \cite{Arndt1997ooa,Gehm2003ule}, we define a length scale corresponding to the inverse wave number of the relative motion $1/\bar{k}_{\rm rel}  = \hbar/(m_r \bar{v}_{\rm rel})$, where $\bar{v}_{\rm rel} = \sqrt{8 k_B T/(\pi m_r)}$ is the mean relative velocity and $m_r$ denotes the reduced mass. The typical interparticle distance sets another length scale, for which we adopt a common definition for two-component Fermi gases, $d = (3\pi^2 n_{\rm tot})^{-1/3}$, where $n_{\rm tot}$ is the total number density of both species in the trap center. For our typical experimental parameters, we obtain $1/\bar{k}_{\rm rel}  \approx 2100\,a_0$ and $d  \approx 2500\,a_0$. The scattering length exceeds $1/\bar{k}_{\rm rel}$ in a magnetic field range of roughly $\pm 0.7\,$G. In this resonant interaction regime, scattering is dominated by the unitarity limitation. In addition to that, the values of $1/\bar{k}_{\rm rel}$ and $d$ are similar, which means that the system is in the crossover between weak and strong interactions. A further length scale is set by the effective range of the resonance \cite{Chin2010fri}. Its value is presently unknown because of the yet undetermined magnetic moment of the molecular state underlying the resonance, but we expect the effective range \cite{SM} to be rather small in comparison to realistic values of the scattering length and the interparticle spacing, so that the interaction physics will be dominated by universal behavior.
 
\begin{figure}
\includegraphics[width=0.95\columnwidth]{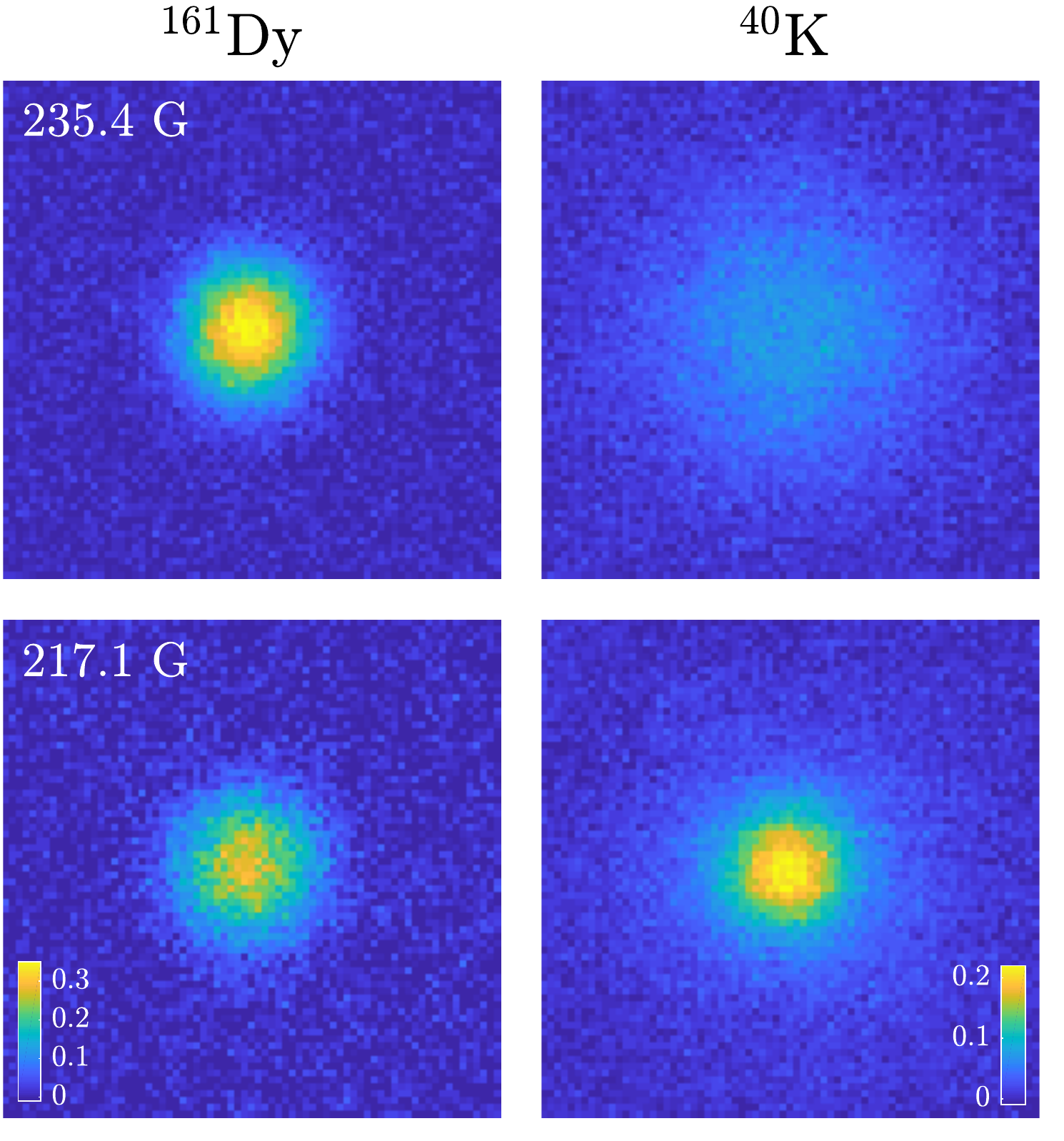}
\caption{Comparison of the expansion of the mixture for weak (upper) and resonant (lower) interspecies interaction. The absorption images show the optical depth for both species (Dy left, K right) after a time of flight of 4.5\,ms. The field of view of all images is $240\,\mu{\rm m} \times 240\,\mu$m.}
\label{Fig2}
\end{figure} 

A striking effect of the resonant interspecies interaction shows up in the expansion of the mixture. In the experiments, the sample was released from the trap right after switching to the target field strength. The absorption images in the upper row of Fig.~\ref{Fig2} illustrate the case of weak interactions ($a \approx -40\,a_0$), realized at $B=235.4$\,G. Here the expansion takes place in a ballistic way and, as expected from the mass ratio, the K component expands much faster than the Dy component. In contrast, in the resonant case (images in the lower row of Fig.~\ref{Fig2}) both components expand with similar sizes. Evidently, the interaction between the two species slows down the expansion of the lighter species and accelerates the expansion of the heavier species. Such a behavior requires many elastic collisions \cite{collrate} on the timescale of the expansion and thus can be interpreted as a hallmark of hydrodynamic behavior.  


\begin{figure}
\includegraphics[width=1\columnwidth]{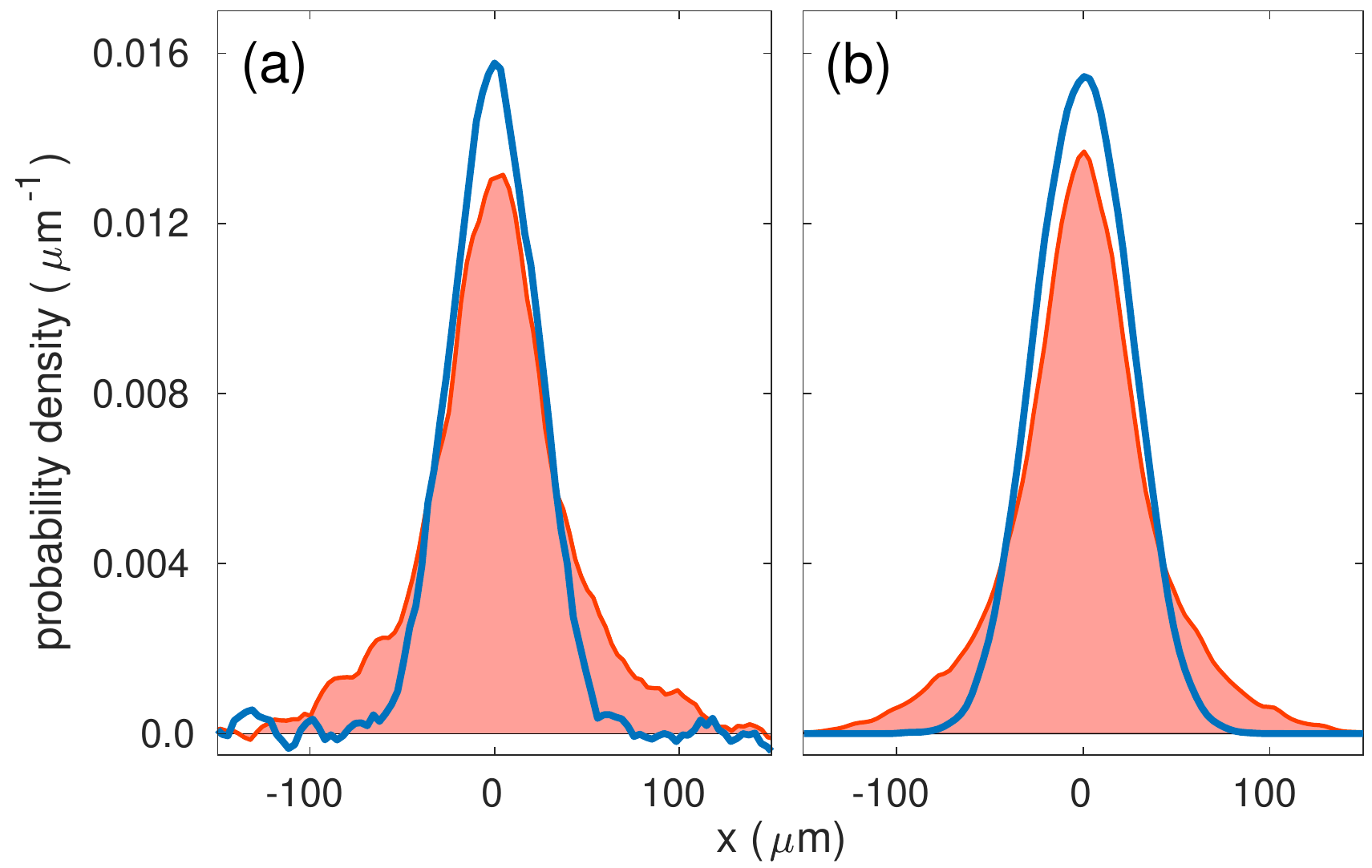}
\caption{Profiles of the hydrodynamically expanding mixture for resonant interaction, (a) experimentally observed and (b)~from a corresponding Monte-Carlo simulation. Shown are the probability densities of doubly-integrated profiles for both Dy (solid blue lines) and K (red curve with filling).}
\label{Fig3}
\end{figure} 

A closer inspection of the spatial profiles of the hydrodynamically expanding mixture reveals an interesting difference between the heavy and the light species; see profiles in Fig.~\ref{Fig3}. While the Dy cloud essentially keeps its near-Gaussian shape, the K cloud (initially about twice smaller than the Dy cloud) develops pronounced side wings. Apparently, the mixture forms a hydrodynamic core surrounded by a larger cloud of ballistically expanding lighter atoms. 

To elucidate the origin of this surprising effect we have carried out a Monte-Carlo simulation \cite{Anderlini2005mfc}, accounting for the classical motion and the quantum-mechanically resonant collisional cross section, which is only limited by the finite relative momentum of a colliding pair \cite{Arndt1997ooa, Gehm2003ule}. For our near-degenerate conditions, we neglect Pauli blocking and interactions beyond two-body collisions. The simulation results in Fig.~\ref{Fig3}(b) reproduce the experimental profiles (a) without any free parameter. The simulation confirms our interpretation in terms of a hydrodynamic core, where both species collide with each other at a large rate, surrounded by a ballistically expanding cloud of light atoms. The physical mechanism for the formation of the latter is the faster diffusion of lighter atoms, which can leak out of the core and, in the absence of the other species, begin to move ballistically. We point out that this bimodality effect is not an experimental imperfection, but a generic feature in the hydrodynamic expansion of a mass-imbalanced mixture.

\begin{figure}
\includegraphics[width=1\columnwidth]{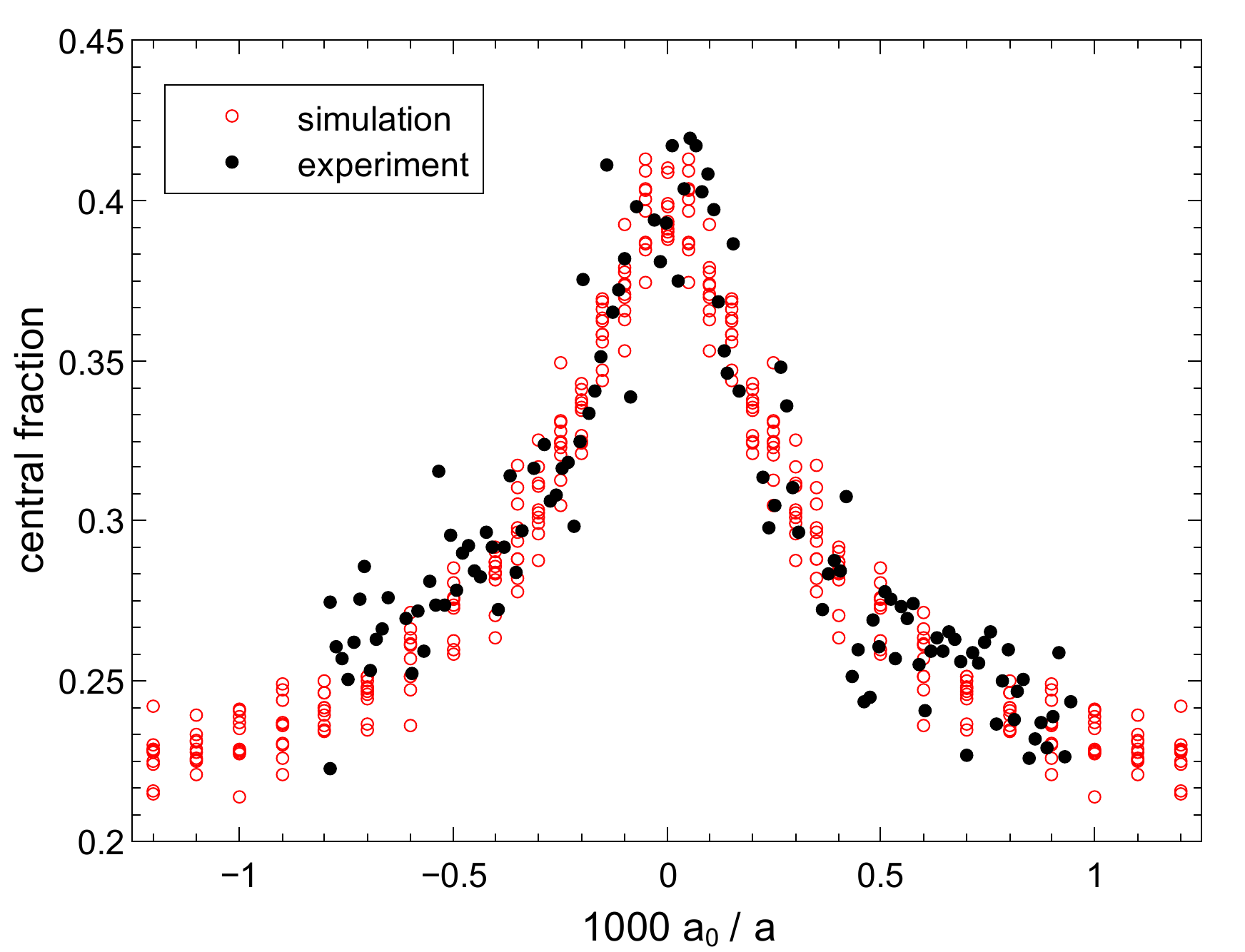}
\caption{Enhancement of the central fraction of K atoms in the expanding mixture. Experimental results for the resonance behavior (filled black symbols) are shown in comparison with Monte-Carlo simulation results (red open symbols).}
\label{Fig4}
\end{figure}  

To investigate the dependence of the hydrodynamic expansion on the scattering length, we recorded two-dimensional expansion profiles (such as in Fig.~\ref{Fig2}) for various values of $B$ in a 2-G wide range around the resonance center. We focus our analysis on the K profiles as they reveal the hydrodynamic core, while the Dy profiles only show a slight increase in width. As a quantitative measure we define the ``central fraction'' as the fraction of K atoms in a circle of particular radius. For the latter we use the $\sqrt{2} \sigma$-width of the non-interacting Dy cloud ($\sim34\,\mu$m at a 4.5-ms time of flight). We find a marked increase of the central fraction from its non-interacting background value 0.22 to a resonant peak value of about 0.40. As a function of the magnetic detuning $B-B_0$, the central fraction shows a pronounced resonance behavior, which closely resembles a Lorentzian curve. From a fit we derive the center $B_0 = 217.04$G and a width (half width at half maximum) of 0.37\,G.  We finally use Eq.~(\ref{eq:a}) with the fixed value $A = 1450$\,G to convert the magnetic detuning scale into an inverse scattering length and plot the data as shown in Fig.~\ref{Fig4}. 

For comparison, we have also employed our Monte-Carlo approach to calculate the central fraction as function of the scattering length. Figure~\ref{Fig4} shows the simulation results (red open symbols) together with the experimental data (black closed symbols). We find that the simulation reproduces the experimental observations very well. This agreement between experiment and theory strongly supports our qualitative and quantitative understanding of both the resonance scenario and the expansion dynamics.

For a precise determination of the resonance center, measurements based on the hydrodynamic expansion can in general provide much sharper resonance features than simple thermalization \cite{Regal2004lom}. While our expansion measurement yielded $217.04(1)$\,G for the resonance center $B_0$, the thermalization measurement \cite{SM} resulted in a value of $217.27(15)$\,G, somewhat higher and with a statistical uncertainty more than an order of magnitude larger. Whether the apparent deviation is a pure statistical effect (about 1.5\,$\sigma$), whether it is caused by magnetic-field control issues \cite{Bfluct}, or whether there are unknown systematic effects behind it requires further investigation. 
We note that anisotropic expansion effects in our nearly spherical trap remain very weak and are barely observable. The anisotropic expansion of a hydrodynamic, strongly interacting Fermi-Fermi system has been studied in our earlier work on a resonant $^6$Li-$^{40}$K mixture \cite{Trenkwalder2011heo}.

\begin{figure}
\includegraphics[width=1\columnwidth]{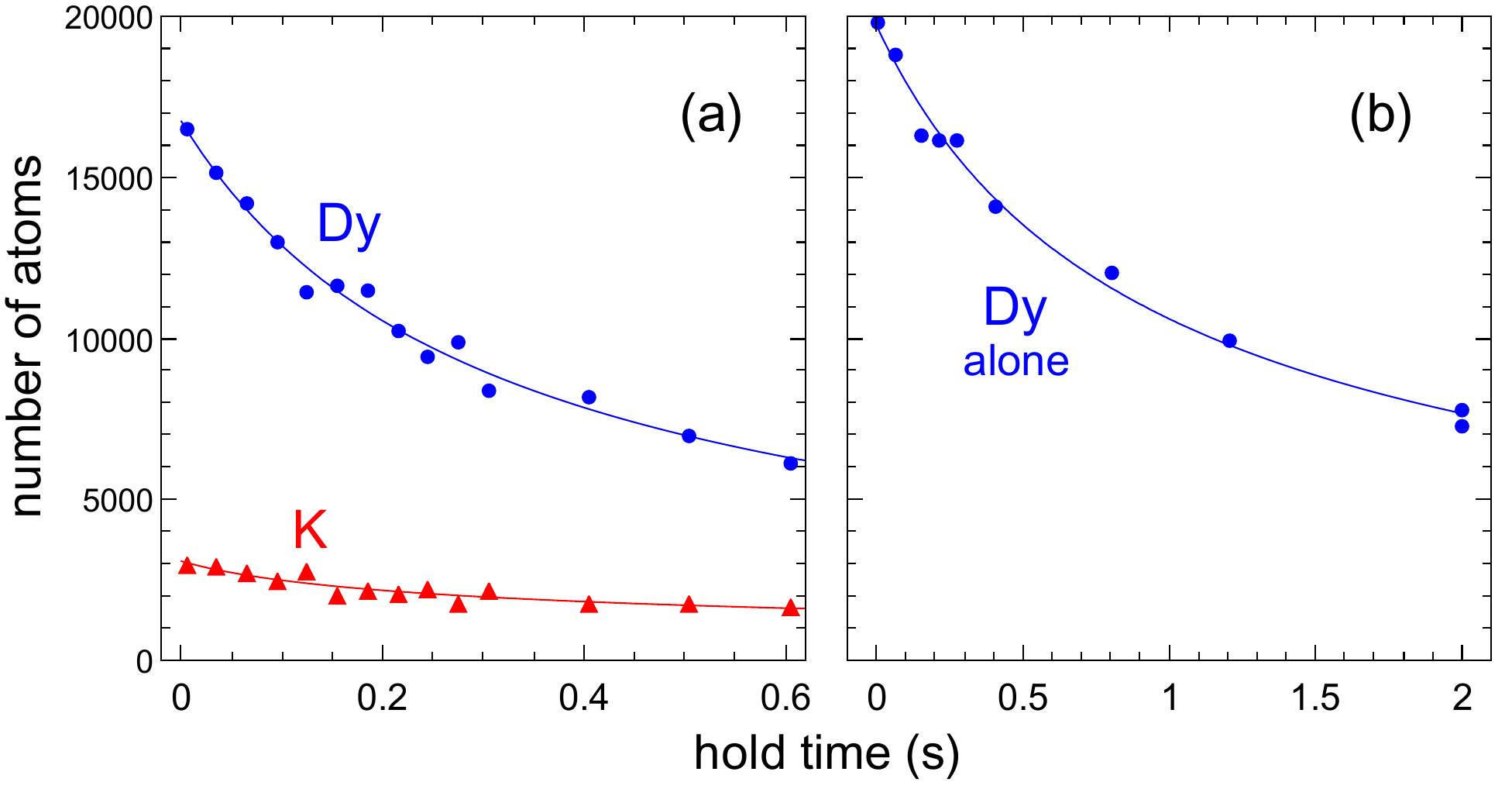}
\caption{Decay of the resonant Dy-K mixture (a) in comparison with a pure Dy sample (b) at magnetic field $B = 217.5$\,G, very close to the resonance pole. The solid lines show fits by a phenomenological model \cite{SM}.}
\label{Fig5}
\end{figure}  

We now turn our attention to the lifetime of the mixture in the resonance region. In general, we find the magnetic-field dependence of losses to exhibit a very complex behavior \cite{SM}. Both Dy-K interspecies and Dy intraspecies losses show strong fluctuations with a variation of the magnetic field. A broad loss feature appears about 0.5\,G below the 217-G resonance, where the scattering length is very large and positive. This feature closely resembles observations made in spin mixtures of $^6$Li \cite{Dieckmann2002doa,Bourdel2003mot,Jochim2004PhD} and $^{40}$K \cite{Regal2004lom}, which have been understood as signature of the formation of weakly bound dimers. In addition to this broad feature, additional narrower structures appear, which make the experiment very sensitive to the particular choice of the magnetic field. Nevertheless, several good regions exist close to the center of the broad Feshbach resonance under conditions, where losses are relatively weak and $s$-wave scattering is deep in the unitarity-limited regime. 

As an example for long lifetimes attainable in the resonance region, Figure~\ref{Fig5} shows a set of measurements taken at field strength of  217.5\,G, for which we estimate a large negative scattering length of $-3000\,a_0$ or even larger. We have fitted and analyzed the decay of the atom numbers following the procedures detailed in \cite{SM}. For the number of K atoms, our data show an initial time constant of about 350\,ms. If we attribute this decay completely to K-Dy-Dy (K-K-Dy) three-body processes, we obtain the upper limits of $4 \times 10^{-25}$\,cm$^6$/s  ($3 \times 10^{-25}$\,cm$^6$/s) for the event rate coefficients. These values are very small compared with other resonant three-body systems that do not involve identical fermions. In Feshbach-resonant Bose-Bose \cite{Barontini2009ooh,Maier2015era,Wacker2016utb} or Bose-Fermi mixtures \cite{Bloom2013tou,Pires2014ooe,Tung2014gso,Ulmanis2016utb,Lous2018pti}, event rate coefficients  have been measured exceeding $10^{-23}$\,cm$^6$/s, i.e.\ at least two orders of magnitude more. In preliminary experiments \cite{TzanovaPhD} on Bose-Fermi mixtures of Dy-K (bosonic  isotope $^{162}$Dy), we have also observed a dramatic increase of resonant three-body losses by orders of magnitude. We attribute the low values of the three-body rate coefficients and thus the long lifetimes in our Fermi-Fermi system to the Pauli suppression of inelastic losses \cite{Petrov2004wbd,Petrov2005spo,Petrov2005dmi}.

The decay of the Dy component in the mixture, displayed in Fig.~\ref{Fig5}(a) by the blue data points and the corresponding fit curve, shows a peculiar behavior. Since we find that about 10 times more Dy atoms are lost as compared to K atoms, three-body interspecies collisions may only explain a small fraction of Dy losses. As Figure~\ref{Fig5}(b) shows, Dy alone exhibits losses even without K being present, but much weaker. Interpreting these losses as Dy intraspecies losses, gives values for the event rate coefficient of $3.4 \times 10^{-25}$\,cm$^6$/s in the presence of K, but only $0.8 \times 10^{-25}$\,cm$^6$/s without K \cite{SM}. These observations point to an unknown mechanism, in which K atoms somehow catalyze the decay of Dy without directly participating in the loss processes. A possible mechanism may be due to elastic collisions with K atoms causing residual evaporation. We tested this in our experiments by recompressing the trap, but did not observe signficant changes in the observed loss behavior. Another hypothesis is based on a spatial contraction (density increase) of the Dy cloud caused by strong interaction effects with K atoms. Considering the zero-temperature limit, we have developed a model \cite{SM} for such an effect, but its applicability is questionable at the temperatures of our present experiments. The explanation of the mysterious enhancement of Dy losses induced by K remains a task for future experiments.

Already our present experiments, carried out near quantum degeneracy ($T/T_F^{\rm K} \approx 0.65$), 
demonstrate that mass imbalance can make a qualitative difference in the physical behavior of a strongly interacting fermion mixture. 
The bimodality observed in the hydrodynamic expansion profile of the lighter component 
is seemingly similar to observations in population-imbalanced spin mixtures near the superfluid phase transition \cite{Zwierlein2006doo}. However, while in the latter case bimodality signals superfluidity, the reason is a different one in our case. Detailed understanding of the expansion dynamics of a Fermi-Fermi mixture in different classical and quantum regimes is thus essential for interpreting the expansion profiles in future work aiming at superfluid regimes.

For reaching lower temperatures and deeper degeneracy, work is in progress to eliminate heating in the transfer from low to high magnetic fields and to implement an additional evaporative cooling stage that takes advantage of the large elastic scattering cross section close to the resonance. The experimental challenge is to realize similar degeneracy conditions near the 217-G resonance as we have achieved at a low magnetic field \cite{Ravensbergen2018poa}.
With some improvements, conditions for superfluid regimes seem to be attainable.  To give an example, a Lifshitz point \cite{Gubbels2009lpi,Baarsma2010pam} in the phase diagram, where zero momentum pairs become unstable, may be expected at a temperature corresponding to about 15\% of the Fermi temperature of the heavy species \cite{lifshitz}. 


In conclusion, we have shown that the $^{161}$Dy-$^{40}$K mixture posseses a broad Feshbach resonance offering favorable conditions for experiments on strongly interacting fermion systems with mass imbalance. In particular, the system features a substantial suppression of inelastic losses near resonance, which is a key requirement for many experiments. Novel interaction regimes, including unconventional superfluid phases, seem to be in reach.

\begin{acknowledgments}
We acknowledge support by the Austrian Science Fund (FWF) within project P32153-N36 and within the Doktoratskolleg ALM (W1259-N27). We thank D. Petrov for enlightening discussions on fermionic suppression effects. We furthermore thank the members of the ultracold atom groups in Innsbruck for many stimulating discussions and for sharing technological know how.
\end{acknowledgments}


\begin{thebibliography}{69}%
\makeatletter
\providecommand \@ifxundefined [1]{%
 \@ifx{#1\undefined}
}%
\providecommand \@ifnum [1]{%
 \ifnum #1\expandafter \@firstoftwo
 \else \expandafter \@secondoftwo
 \fi
}%
\providecommand \@ifx [1]{%
 \ifx #1\expandafter \@firstoftwo
 \else \expandafter \@secondoftwo
 \fi
}%
\providecommand \natexlab [1]{#1}%
\providecommand \enquote  [1]{``#1''}%
\providecommand \bibnamefont  [1]{#1}%
\providecommand \bibfnamefont [1]{#1}%
\providecommand \citenamefont [1]{#1}%
\providecommand \href@noop [0]{\@secondoftwo}%
\providecommand \href [0]{\begingroup \@sanitize@url \@href}%
\providecommand \@href[1]{\@@startlink{#1}\@@href}%
\providecommand \@@href[1]{\endgroup#1\@@endlink}%
\providecommand \@sanitize@url [0]{\catcode `\\12\catcode `\$12\catcode
  `\&12\catcode `\#12\catcode `\^12\catcode `\_12\catcode `\%12\relax}%
\providecommand \@@startlink[1]{}%
\providecommand \@@endlink[0]{}%
\providecommand \url  [0]{\begingroup\@sanitize@url \@url }%
\providecommand \@url [1]{\endgroup\@href {#1}{\urlprefix }}%
\providecommand \urlprefix  [0]{URL }%
\providecommand \Eprint [0]{\href }%
\providecommand \doibase [0]{http://dx.doi.org/}%
\providecommand \selectlanguage [0]{\@gobble}%
\providecommand \bibinfo  [0]{\@secondoftwo}%
\providecommand \bibfield  [0]{\@secondoftwo}%
\providecommand \translation [1]{[#1]}%
\providecommand \BibitemOpen [0]{}%
\providecommand \bibitemStop [0]{}%
\providecommand \bibitemNoStop [0]{.\EOS\space}%
\providecommand \EOS [0]{\spacefactor3000\relax}%
\providecommand \BibitemShut  [1]{\csname bibitem#1\endcsname}%
\let\auto@bib@innerbib\@empty
\bibitem [{\citenamefont {Inguscio}\ \emph {et~al.}(2008)\citenamefont
  {Inguscio}, \citenamefont {Ketterle},\ and\ \citenamefont
  {Salomon}}]{Inguscio2006ufg}%
  \BibitemOpen
  \bibinfo {editor} {\bibfnamefont {M.}~\bibnamefont {Inguscio}}, \bibinfo
  {editor} {\bibfnamefont {W.}~\bibnamefont {Ketterle}}, \ and\ \bibinfo
  {editor} {\bibfnamefont {C.}~\bibnamefont {Salomon}},\ eds.,\ \href@noop {}
  {\emph {\bibinfo {title} {Ultra-cold Fermi Gases}}}\ (\bibinfo  {publisher}
  {IOS Press, Amsterdam},\ \bibinfo {year} {2008})\ \bibinfo {note}
  {{P}roceedings of the International School of Physics ``Enrico Fermi'',
  Course CLXIV, Varenna, 20-30 June 2006}\BibitemShut {NoStop}%
\bibitem [{\citenamefont {Pitaevskii}\ and\ \citenamefont
  {Stringari}(2016)}]{Pitaevskii2016book}%
  \BibitemOpen
  \bibfield  {author} {\bibinfo {author} {\bibfnamefont {L.}~\bibnamefont
  {Pitaevskii}}\ and\ \bibinfo {author} {\bibfnamefont {S.}~\bibnamefont
  {Stringari}},\ }\href@noop {} {\emph {\bibinfo {title} {Bose-Einstein
  Condensation and Superfluidity}}}\ (\bibinfo  {publisher} {Oxford University
  Press},\ \bibinfo {year} {2016})\BibitemShut {NoStop}%
\bibitem [{\citenamefont {Zwerger}(2012)}]{Zwerger2012tbb}%
  \BibitemOpen
  \bibinfo {editor} {\bibfnamefont {W.}~\bibnamefont {Zwerger}},\ ed.,\
  \href@noop {} {\emph {\bibinfo {title} {The BCS-BEC Crossover and the Unitary
  Fermi Gas}}}\ (\bibinfo  {publisher} {Springer, Berlin Heidelberg},\ \bibinfo
  {year} {2012})\BibitemShut {NoStop}%
\bibitem [{\citenamefont {Strinati}\ \emph {et~al.}(2018)\citenamefont
  {Strinati}, \citenamefont {Pieri}, \citenamefont {R{\"o}pke}, \citenamefont
  {Schuck},\ and\ \citenamefont {Urban}}]{Strinati2018tbb}%
  \BibitemOpen
  \bibfield  {author} {\bibinfo {author} {\bibfnamefont {G.~C.}\ \bibnamefont
  {Strinati}}, \bibinfo {author} {\bibfnamefont {P.}~\bibnamefont {Pieri}},
  \bibinfo {author} {\bibfnamefont {G.}~\bibnamefont {R{\"o}pke}}, \bibinfo
  {author} {\bibfnamefont {P.}~\bibnamefont {Schuck}}, \ and\ \bibinfo {author}
  {\bibfnamefont {M.}~\bibnamefont {Urban}},\ }\href {\doibase
  10.1016/j.physrep.2018.02.004} {\bibfield  {journal} {\bibinfo  {journal}
  {Phys. Rep.}\ }\textbf {\bibinfo {volume} {738}},\ \bibinfo {pages} {1}
  (\bibinfo {year} {2018})}\BibitemShut {NoStop}%
\bibitem [{\citenamefont {Bennemann}\ and\ \citenamefont
  {Ketterson}(2014)}]{Bennemann2014both}%
  \BibitemOpen
  \bibfield  {author} {\bibinfo {author} {\bibfnamefont {K.-H.}\ \bibnamefont
  {Bennemann}}\ and\ \bibinfo {author} {\bibfnamefont {J.~B.}\ \bibnamefont
  {Ketterson}},\ }\href@noop {} {\emph {\bibinfo {title} {Novel Superfluids:
  Volumes 1 and 2}}}\ (\bibinfo  {publisher} {Oxford University Press,
  Oxford},\ \bibinfo {year} {2013, 2014})\BibitemShut {NoStop}%
\bibitem [{\citenamefont {Chin}\ \emph {et~al.}(2010)\citenamefont {Chin},
  \citenamefont {Grimm}, \citenamefont {Julienne},\ and\ \citenamefont
  {Tiesinga}}]{Chin2010fri}%
  \BibitemOpen
  \bibfield  {author} {\bibinfo {author} {\bibfnamefont {C.}~\bibnamefont
  {Chin}}, \bibinfo {author} {\bibfnamefont {R.}~\bibnamefont {Grimm}},
  \bibinfo {author} {\bibfnamefont {P.~S.}\ \bibnamefont {Julienne}}, \ and\
  \bibinfo {author} {\bibfnamefont {E.}~\bibnamefont {Tiesinga}},\ }\href
  {\doibase doi.org/10.1103/RevModPhys.82.1225} {\bibfield  {journal} {\bibinfo
   {journal} {Rev. Mod. Phys.}\ }\textbf {\bibinfo {volume} {82}},\ \bibinfo
  {pages} {1225} (\bibinfo {year} {2010})}\BibitemShut {NoStop}%
\bibitem [{\citenamefont {Gubbels}\ and\ \citenamefont
  {Stoof}(2013)}]{Gubbels2013ifg}%
  \BibitemOpen
  \bibfield  {author} {\bibinfo {author} {\bibfnamefont {K.}~\bibnamefont
  {Gubbels}}\ and\ \bibinfo {author} {\bibfnamefont {H.}~\bibnamefont
  {Stoof}},\ }\href {\doibase 10.1016/j.physrep.2012.11.004} {\bibfield
  {journal} {\bibinfo  {journal} {Phys. Rep.}\ }\textbf {\bibinfo {volume}
  {525}},\ \bibinfo {pages} {255 } (\bibinfo {year} {2013})}\BibitemShut
  {NoStop}%
\bibitem [{\citenamefont {Gubbels}\ \emph {et~al.}(2009)\citenamefont
  {Gubbels}, \citenamefont {Baarsma},\ and\ \citenamefont
  {Stoof}}]{Gubbels2009lpi}%
  \BibitemOpen
  \bibfield  {author} {\bibinfo {author} {\bibfnamefont {K.~B.}\ \bibnamefont
  {Gubbels}}, \bibinfo {author} {\bibfnamefont {J.~E.}\ \bibnamefont
  {Baarsma}}, \ and\ \bibinfo {author} {\bibfnamefont {H.~T.~C.}\ \bibnamefont
  {Stoof}},\ }\href {\doibase 10.1103/PhysRevLett.103.195301} {\bibfield
  {journal} {\bibinfo  {journal} {Phys. Rev. Lett.}\ }\textbf {\bibinfo
  {volume} {103}},\ \bibinfo {pages} {195301} (\bibinfo {year}
  {2009})}\BibitemShut {NoStop}%
\bibitem [{\citenamefont {Wang}\ \emph {et~al.}(2017)\citenamefont {Wang},
  \citenamefont {Che}, \citenamefont {Zhang},\ and\ \citenamefont
  {Chen}}]{Wang2017eeo}%
  \BibitemOpen
  \bibfield  {author} {\bibinfo {author} {\bibfnamefont {J.}~\bibnamefont
  {Wang}}, \bibinfo {author} {\bibfnamefont {Y.}~\bibnamefont {Che}}, \bibinfo
  {author} {\bibfnamefont {L.}~\bibnamefont {Zhang}}, \ and\ \bibinfo {author}
  {\bibfnamefont {Q.}~\bibnamefont {Chen}},\ }\href {\doibase
  10.1038/srep39783} {\bibfield  {journal} {\bibinfo  {journal} {Sci. Rep.}\
  }\textbf {\bibinfo {volume} {7}},\ \bibinfo {pages} {39783} (\bibinfo {year}
  {2017})}\BibitemShut {NoStop}%
\bibitem [{\citenamefont {Fulde}\ and\ \citenamefont
  {Ferrell}(1964)}]{Fulde1964sia}%
  \BibitemOpen
  \bibfield  {author} {\bibinfo {author} {\bibfnamefont {P.}~\bibnamefont
  {Fulde}}\ and\ \bibinfo {author} {\bibfnamefont {R.~A.}\ \bibnamefont
  {Ferrell}},\ }\href {\doibase 10.1103/PhysRev.135.A550} {\bibfield  {journal}
  {\bibinfo  {journal} {Phys. Rev.}\ }\textbf {\bibinfo {volume} {135}},\
  \bibinfo {pages} {A550} (\bibinfo {year} {1964})}\BibitemShut {NoStop}%
\bibitem [{\citenamefont {Larkin}\ and\ \citenamefont
  {Ovchinnikov}(1965)}]{Larkin1964iso}%
  \BibitemOpen
  \bibfield  {author} {\bibinfo {author} {\bibfnamefont {A.~I.}\ \bibnamefont
  {Larkin}}\ and\ \bibinfo {author} {\bibfnamefont {Y.~N.}\ \bibnamefont
  {Ovchinnikov}},\ }\href@noop {} {\bibfield  {journal} {\bibinfo  {journal}
  {Sov. Phys. JETP}\ }\textbf {\bibinfo {volume} {20}},\ \bibinfo {pages} {762}
  (\bibinfo {year} {1965})}\BibitemShut {NoStop}%
\bibitem [{\citenamefont {Radzihovsky}\ and\ \citenamefont
  {Sheehy}(2010)}]{radzihovsky2010ifr}%
  \BibitemOpen
  \bibfield  {author} {\bibinfo {author} {\bibfnamefont {L.}~\bibnamefont
  {Radzihovsky}}\ and\ \bibinfo {author} {\bibfnamefont {D.~E.}\ \bibnamefont
  {Sheehy}},\ }\href {\doibase 10.1088/0034-4885/73/7/076501} {\bibfield
  {journal} {\bibinfo  {journal} {Rep. Prog. Phys.}\ }\textbf {\bibinfo
  {volume} {73}},\ \bibinfo {pages} {076501} (\bibinfo {year}
  {2010})}\BibitemShut {NoStop}%
\bibitem [{\citenamefont {Petrov}\ \emph {et~al.}(2004)\citenamefont {Petrov},
  \citenamefont {Salomon},\ and\ \citenamefont {Shlyapnikov}}]{Petrov2004wbd}%
  \BibitemOpen
  \bibfield  {author} {\bibinfo {author} {\bibfnamefont {D.~S.}\ \bibnamefont
  {Petrov}}, \bibinfo {author} {\bibfnamefont {C.}~\bibnamefont {Salomon}}, \
  and\ \bibinfo {author} {\bibfnamefont {G.~V.}\ \bibnamefont {Shlyapnikov}},\
  }\href {\doibase 10.1103/PhysRevLett.93.090404} {\bibfield  {journal}
  {\bibinfo  {journal} {Phys. Rev. Lett.}\ }\textbf {\bibinfo {volume} {93}},\
  \bibinfo {pages} {090404} (\bibinfo {year} {2004})}\BibitemShut {NoStop}%
\bibitem [{\citenamefont {Petrov}\ \emph
  {et~al.}(2005{\natexlab{a}})\citenamefont {Petrov}, \citenamefont {Salomon},\
  and\ \citenamefont {Shlyapnikov}}]{Petrov2005spo}%
  \BibitemOpen
  \bibfield  {author} {\bibinfo {author} {\bibfnamefont {D.~S.}\ \bibnamefont
  {Petrov}}, \bibinfo {author} {\bibfnamefont {C.}~\bibnamefont {Salomon}}, \
  and\ \bibinfo {author} {\bibfnamefont {G.~V.}\ \bibnamefont {Shlyapnikov}},\
  }\href {\doibase 10.1103/PhysRevA.71.012708} {\bibfield  {journal} {\bibinfo
  {journal} {Phys. Rev. A}\ }\textbf {\bibinfo {volume} {71}},\ \bibinfo {eid}
  {012708} (\bibinfo {year} {2005}{\natexlab{a}})}\BibitemShut {NoStop}%
\bibitem [{\citenamefont {Levinsen}\ and\ \citenamefont
  {Petrov}(2011)}]{Levinsen2011ada}%
  \BibitemOpen
  \bibfield  {author} {\bibinfo {author} {\bibfnamefont {J.}~\bibnamefont
  {Levinsen}}\ and\ \bibinfo {author} {\bibfnamefont {D.}~\bibnamefont
  {Petrov}},\ }\href {\doibase 10.1140/epjd/e2011-20071-x} {\bibfield
  {journal} {\bibinfo  {journal} {Eur. Phys. J. D}\ }\textbf {\bibinfo {volume}
  {65}},\ \bibinfo {pages} {67} (\bibinfo {year} {2011})}\BibitemShut {NoStop}%
\bibitem [{\citenamefont {Petrov}\ \emph
  {et~al.}(2005{\natexlab{b}})\citenamefont {Petrov}, \citenamefont {Salomon},\
  and\ \citenamefont {Shlyapnikov}}]{Petrov2005dmi}%
  \BibitemOpen
  \bibfield  {author} {\bibinfo {author} {\bibfnamefont {D.~S.}\ \bibnamefont
  {Petrov}}, \bibinfo {author} {\bibfnamefont {C.}~\bibnamefont {Salomon}}, \
  and\ \bibinfo {author} {\bibfnamefont {G.~V.}\ \bibnamefont {Shlyapnikov}},\
  }\href {\doibase 10.1088/0953-4075/38/9/014} {\bibfield  {journal} {\bibinfo
  {journal} {J. Phys. B}\ }\textbf {\bibinfo {volume} {38}},\ \bibinfo {pages}
  {S645} (\bibinfo {year} {2005}{\natexlab{b}})}\BibitemShut {NoStop}%
\bibitem [{\citenamefont {Trenkwalder}\ \emph {et~al.}(2011)\citenamefont
  {Trenkwalder}, \citenamefont {Kohstall}, \citenamefont {Zaccanti},
  \citenamefont {Naik}, \citenamefont {Sidorov}, \citenamefont {Schreck},\ and\
  \citenamefont {Grimm}}]{Trenkwalder2011heo}%
  \BibitemOpen
  \bibfield  {author} {\bibinfo {author} {\bibfnamefont {A.}~\bibnamefont
  {Trenkwalder}}, \bibinfo {author} {\bibfnamefont {C.}~\bibnamefont
  {Kohstall}}, \bibinfo {author} {\bibfnamefont {M.}~\bibnamefont {Zaccanti}},
  \bibinfo {author} {\bibfnamefont {D.}~\bibnamefont {Naik}}, \bibinfo {author}
  {\bibfnamefont {A.~I.}\ \bibnamefont {Sidorov}}, \bibinfo {author}
  {\bibfnamefont {F.}~\bibnamefont {Schreck}}, \ and\ \bibinfo {author}
  {\bibfnamefont {R.}~\bibnamefont {Grimm}},\ }\href {\doibase
  10.1103/PhysRevLett.106.115304} {\bibfield  {journal} {\bibinfo  {journal}
  {Phys. Rev. Lett.}\ }\textbf {\bibinfo {volume} {106}},\ \bibinfo {pages}
  {115304} (\bibinfo {year} {2011})}\BibitemShut {NoStop}%
\bibitem [{\citenamefont {Jag}\ \emph {et~al.}(2014)\citenamefont {Jag},
  \citenamefont {Zaccanti}, \citenamefont {Cetina}, \citenamefont {Lous},
  \citenamefont {Schreck}, \citenamefont {Grimm}, \citenamefont {Petrov},\ and\
  \citenamefont {Levinsen}}]{Jag2014ooa}%
  \BibitemOpen
  \bibfield  {author} {\bibinfo {author} {\bibfnamefont {M.}~\bibnamefont
  {Jag}}, \bibinfo {author} {\bibfnamefont {M.}~\bibnamefont {Zaccanti}},
  \bibinfo {author} {\bibfnamefont {M.}~\bibnamefont {Cetina}}, \bibinfo
  {author} {\bibfnamefont {R.~S.}\ \bibnamefont {Lous}}, \bibinfo {author}
  {\bibfnamefont {F.}~\bibnamefont {Schreck}}, \bibinfo {author} {\bibfnamefont
  {R.}~\bibnamefont {Grimm}}, \bibinfo {author} {\bibfnamefont {D.~S.}\
  \bibnamefont {Petrov}}, \ and\ \bibinfo {author} {\bibfnamefont
  {J.}~\bibnamefont {Levinsen}},\ }\href {\doibase
  10.1103/PhysRevLett.112.075302} {\bibfield  {journal} {\bibinfo  {journal}
  {Phys. Rev. Lett.}\ }\textbf {\bibinfo {volume} {112}},\ \bibinfo {pages}
  {075302} (\bibinfo {year} {2014})}\BibitemShut {NoStop}%
\bibitem [{\citenamefont {Wille}\ \emph {et~al.}(2008)\citenamefont {Wille},
  \citenamefont {Spiegelhalder}, \citenamefont {Kerner}, \citenamefont {Naik},
  \citenamefont {Trenkwalder}, \citenamefont {Hendl}, \citenamefont {Schreck},
  \citenamefont {Grimm}, \citenamefont {Tiecke}, \citenamefont {Walraven},
  \citenamefont {Kokkelmans}, \citenamefont {Tiesinga},\ and\ \citenamefont
  {Julienne}}]{Wille2008eau}%
  \BibitemOpen
  \bibfield  {author} {\bibinfo {author} {\bibfnamefont {E.}~\bibnamefont
  {Wille}}, \bibinfo {author} {\bibfnamefont {F.~M.}\ \bibnamefont
  {Spiegelhalder}}, \bibinfo {author} {\bibfnamefont {G.}~\bibnamefont
  {Kerner}}, \bibinfo {author} {\bibfnamefont {D.}~\bibnamefont {Naik}},
  \bibinfo {author} {\bibfnamefont {A.}~\bibnamefont {Trenkwalder}}, \bibinfo
  {author} {\bibfnamefont {G.}~\bibnamefont {Hendl}}, \bibinfo {author}
  {\bibfnamefont {F.}~\bibnamefont {Schreck}}, \bibinfo {author} {\bibfnamefont
  {R.}~\bibnamefont {Grimm}}, \bibinfo {author} {\bibfnamefont {T.~G.}\
  \bibnamefont {Tiecke}}, \bibinfo {author} {\bibfnamefont {J.~T.~M.}\
  \bibnamefont {Walraven}}, \bibinfo {author} {\bibfnamefont {S.~J. J. M.~F.}\
  \bibnamefont {Kokkelmans}}, \bibinfo {author} {\bibfnamefont
  {E.}~\bibnamefont {Tiesinga}}, \ and\ \bibinfo {author} {\bibfnamefont
  {P.~S.}\ \bibnamefont {Julienne}},\ }\href {\doibase
  10.1103/PhysRevLett.100.053201} {\bibfield  {journal} {\bibinfo  {journal}
  {Phys. Rev. Lett.}\ }\textbf {\bibinfo {volume} {100}},\ \bibinfo {eid}
  {053201} (\bibinfo {year} {2008})}\BibitemShut {NoStop}%
\bibitem [{\citenamefont {Tiecke}\ \emph {et~al.}(2010)\citenamefont {Tiecke},
  \citenamefont {Goosen}, \citenamefont {Ludewig}, \citenamefont {Gensemer},
  \citenamefont {Kraft}, \citenamefont {Kokkelmans},\ and\ \citenamefont
  {Walraven}}]{Tiecke2010bfr}%
  \BibitemOpen
  \bibfield  {author} {\bibinfo {author} {\bibfnamefont {T.~G.}\ \bibnamefont
  {Tiecke}}, \bibinfo {author} {\bibfnamefont {M.~R.}\ \bibnamefont {Goosen}},
  \bibinfo {author} {\bibfnamefont {A.}~\bibnamefont {Ludewig}}, \bibinfo
  {author} {\bibfnamefont {S.~D.}\ \bibnamefont {Gensemer}}, \bibinfo {author}
  {\bibfnamefont {S.}~\bibnamefont {Kraft}}, \bibinfo {author} {\bibfnamefont
  {S.~J. J. M.~F.}\ \bibnamefont {Kokkelmans}}, \ and\ \bibinfo {author}
  {\bibfnamefont {J.~T.~M.}\ \bibnamefont {Walraven}},\ }\href {\doibase
  10.1103/PhysRevLett.104.053202} {\bibfield  {journal} {\bibinfo  {journal}
  {Phys. Rev. Lett.}\ }\textbf {\bibinfo {volume} {104}},\ \bibinfo {pages}
  {053202} (\bibinfo {year} {2010})}\BibitemShut {NoStop}%
\bibitem [{\citenamefont {Naik}\ \emph {et~al.}(2011)\citenamefont {Naik},
  \citenamefont {Trenkwalder}, \citenamefont {Kohstall}, \citenamefont
  {Spiegelhalder}, \citenamefont {Zaccanti}, \citenamefont {Hendl},
  \citenamefont {Schreck}, \citenamefont {Grimm}, \citenamefont {Hanna},\ and\
  \citenamefont {Julienne}}]{Naik2011fri}%
  \BibitemOpen
  \bibfield  {author} {\bibinfo {author} {\bibfnamefont {D.}~\bibnamefont
  {Naik}}, \bibinfo {author} {\bibfnamefont {A.}~\bibnamefont {Trenkwalder}},
  \bibinfo {author} {\bibfnamefont {C.}~\bibnamefont {Kohstall}}, \bibinfo
  {author} {\bibfnamefont {F.~M.}\ \bibnamefont {Spiegelhalder}}, \bibinfo
  {author} {\bibfnamefont {M.}~\bibnamefont {Zaccanti}}, \bibinfo {author}
  {\bibfnamefont {G.}~\bibnamefont {Hendl}}, \bibinfo {author} {\bibfnamefont
  {F.}~\bibnamefont {Schreck}}, \bibinfo {author} {\bibfnamefont
  {R.}~\bibnamefont {Grimm}}, \bibinfo {author} {\bibfnamefont
  {T.}~\bibnamefont {Hanna}}, \ and\ \bibinfo {author} {\bibfnamefont
  {P.}~\bibnamefont {Julienne}},\ }\href {\doibase 10.1140/epjd/e2010-10591-2}
  {\bibfield  {journal} {\bibinfo  {journal} {Eur. Phys. J. D}\ }\textbf
  {\bibinfo {volume} {65}},\ \bibinfo {pages} {55} (\bibinfo {year}
  {2011})}\BibitemShut {NoStop}%
\bibitem [{\citenamefont {Jag}\ \emph {et~al.}(2016)\citenamefont {Jag},
  \citenamefont {Cetina}, \citenamefont {Lous}, \citenamefont {Grimm},
  \citenamefont {Levinsen},\ and\ \citenamefont {Petrov}}]{Jag2016lof}%
  \BibitemOpen
  \bibfield  {author} {\bibinfo {author} {\bibfnamefont {M.}~\bibnamefont
  {Jag}}, \bibinfo {author} {\bibfnamefont {M.}~\bibnamefont {Cetina}},
  \bibinfo {author} {\bibfnamefont {R.~S.}\ \bibnamefont {Lous}}, \bibinfo
  {author} {\bibfnamefont {R.}~\bibnamefont {Grimm}}, \bibinfo {author}
  {\bibfnamefont {J.}~\bibnamefont {Levinsen}}, \ and\ \bibinfo {author}
  {\bibfnamefont {D.~S.}\ \bibnamefont {Petrov}},\ }\href {\doibase
  10.1103/PhysRevA.94.062706} {\bibfield  {journal} {\bibinfo  {journal} {Phys.
  Rev. A}\ }\textbf {\bibinfo {volume} {94}},\ \bibinfo {pages} {062706}
  (\bibinfo {year} {2016})}\BibitemShut {NoStop}%
\bibitem [{\citenamefont {Lu}\ \emph {et~al.}(2011)\citenamefont {Lu},
  \citenamefont {Burdick}, \citenamefont {Youn},\ and\ \citenamefont
  {Lev}}]{Lu2011sdb}%
  \BibitemOpen
  \bibfield  {author} {\bibinfo {author} {\bibfnamefont {M.}~\bibnamefont
  {Lu}}, \bibinfo {author} {\bibfnamefont {N.~Q.}\ \bibnamefont {Burdick}},
  \bibinfo {author} {\bibfnamefont {S.~H.}\ \bibnamefont {Youn}}, \ and\
  \bibinfo {author} {\bibfnamefont {B.~L.}\ \bibnamefont {Lev}},\ }\href
  {\doibase 10.1103/PhysRevLett.107.190401} {\bibfield  {journal} {\bibinfo
  {journal} {Phys. Rev. Lett.}\ }\textbf {\bibinfo {volume} {107}},\ \bibinfo
  {pages} {190401} (\bibinfo {year} {2011})}\BibitemShut {NoStop}%
\bibitem [{\citenamefont {Aikawa}\ \emph {et~al.}(2012)\citenamefont {Aikawa},
  \citenamefont {Frisch}, \citenamefont {Mark}, \citenamefont {Baier},
  \citenamefont {Rietzler}, \citenamefont {Grimm},\ and\ \citenamefont
  {Ferlaino}}]{Aikawa2012bec}%
  \BibitemOpen
  \bibfield  {author} {\bibinfo {author} {\bibfnamefont {K.}~\bibnamefont
  {Aikawa}}, \bibinfo {author} {\bibfnamefont {A.}~\bibnamefont {Frisch}},
  \bibinfo {author} {\bibfnamefont {M.}~\bibnamefont {Mark}}, \bibinfo {author}
  {\bibfnamefont {S.}~\bibnamefont {Baier}}, \bibinfo {author} {\bibfnamefont
  {A.}~\bibnamefont {Rietzler}}, \bibinfo {author} {\bibfnamefont
  {R.}~\bibnamefont {Grimm}}, \ and\ \bibinfo {author} {\bibfnamefont
  {F.}~\bibnamefont {Ferlaino}},\ }\href {\doibase
  10.1103/PhysRevLett.108.210401} {\bibfield  {journal} {\bibinfo  {journal}
  {Phys. Rev. Lett.}\ }\textbf {\bibinfo {volume} {108}},\ \bibinfo {pages}
  {210401} (\bibinfo {year} {2012})}\BibitemShut {NoStop}%
\bibitem [{\citenamefont {{Lu}}\ \emph {et~al.}(2012)\citenamefont {{Lu}},
  \citenamefont {{Burdick}},\ and\ \citenamefont {{Lev}}}]{Lu2012qdd}%
  \BibitemOpen
  \bibfield  {author} {\bibinfo {author} {\bibfnamefont {M.}~\bibnamefont
  {{Lu}}}, \bibinfo {author} {\bibfnamefont {N.~Q.}\ \bibnamefont {{Burdick}}},
  \ and\ \bibinfo {author} {\bibfnamefont {B.~L.}\ \bibnamefont {{Lev}}},\
  }\href {\doibase 10.1103/PhysRevLett.108.215301} {\bibfield  {journal}
  {\bibinfo  {journal} {Phys. Rev. Lett.}\ }\textbf {\bibinfo {volume} {108}},\
  \bibinfo {pages} {215301} (\bibinfo {year} {2012})}\BibitemShut {NoStop}%
\bibitem [{\citenamefont {Aikawa}\ \emph {et~al.}(2014)\citenamefont {Aikawa},
  \citenamefont {Frisch}, \citenamefont {Mark}, \citenamefont {Baier},
  \citenamefont {Grimm},\ and\ \citenamefont {Ferlaino}}]{Aikawa2014rfd}%
  \BibitemOpen
  \bibfield  {author} {\bibinfo {author} {\bibfnamefont {K.}~\bibnamefont
  {Aikawa}}, \bibinfo {author} {\bibfnamefont {A.}~\bibnamefont {Frisch}},
  \bibinfo {author} {\bibfnamefont {M.}~\bibnamefont {Mark}}, \bibinfo {author}
  {\bibfnamefont {S.}~\bibnamefont {Baier}}, \bibinfo {author} {\bibfnamefont
  {R.}~\bibnamefont {Grimm}}, \ and\ \bibinfo {author} {\bibfnamefont
  {F.}~\bibnamefont {Ferlaino}},\ }\href {\doibase
  10.1103/PhysRevLett.112.010404} {\bibfield  {journal} {\bibinfo  {journal}
  {Phys. Rev. Lett.}\ }\textbf {\bibinfo {volume} {112}},\ \bibinfo {pages}
  {010404} (\bibinfo {year} {2014})}\BibitemShut {NoStop}%
\bibitem [{\citenamefont {Trautmann}\ \emph {et~al.}(2018)\citenamefont
  {Trautmann}, \citenamefont {Ilzh\"ofer}, \citenamefont {Durastante},
  \citenamefont {Politi}, \citenamefont {Sohmen}, \citenamefont {Mark},\ and\
  \citenamefont {Ferlaino}}]{Trautmann2018dqm}%
  \BibitemOpen
  \bibfield  {author} {\bibinfo {author} {\bibfnamefont {A.}~\bibnamefont
  {Trautmann}}, \bibinfo {author} {\bibfnamefont {P.}~\bibnamefont
  {Ilzh\"ofer}}, \bibinfo {author} {\bibfnamefont {G.}~\bibnamefont
  {Durastante}}, \bibinfo {author} {\bibfnamefont {C.}~\bibnamefont {Politi}},
  \bibinfo {author} {\bibfnamefont {M.}~\bibnamefont {Sohmen}}, \bibinfo
  {author} {\bibfnamefont {M.~J.}\ \bibnamefont {Mark}}, \ and\ \bibinfo
  {author} {\bibfnamefont {F.}~\bibnamefont {Ferlaino}},\ }\href {\doibase
  10.1103/PhysRevLett.121.213601} {\bibfield  {journal} {\bibinfo  {journal}
  {Phys. Rev. Lett.}\ }\textbf {\bibinfo {volume} {121}},\ \bibinfo {pages}
  {213601} (\bibinfo {year} {2018})}\BibitemShut {NoStop}%
\bibitem [{\citenamefont {Baier}\ \emph {et~al.}(2018)\citenamefont {Baier},
  \citenamefont {Petter}, \citenamefont {Becher}, \citenamefont {Patscheider},
  \citenamefont {Natale}, \citenamefont {Chomaz}, \citenamefont {Mark},\ and\
  \citenamefont {Ferlaino}}]{Baier2018roa}%
  \BibitemOpen
  \bibfield  {author} {\bibinfo {author} {\bibfnamefont {S.}~\bibnamefont
  {Baier}}, \bibinfo {author} {\bibfnamefont {D.}~\bibnamefont {Petter}},
  \bibinfo {author} {\bibfnamefont {J.~H.}\ \bibnamefont {Becher}}, \bibinfo
  {author} {\bibfnamefont {A.}~\bibnamefont {Patscheider}}, \bibinfo {author}
  {\bibfnamefont {G.}~\bibnamefont {Natale}}, \bibinfo {author} {\bibfnamefont
  {L.}~\bibnamefont {Chomaz}}, \bibinfo {author} {\bibfnamefont {M.~J.}\
  \bibnamefont {Mark}}, \ and\ \bibinfo {author} {\bibfnamefont
  {F.}~\bibnamefont {Ferlaino}},\ }\href {\doibase
  10.1103/PhysRevLett.121.093602} {\bibfield  {journal} {\bibinfo  {journal}
  {Phys. Rev. Lett.}\ }\textbf {\bibinfo {volume} {121}},\ \bibinfo {pages}
  {093602} (\bibinfo {year} {2018})}\BibitemShut {NoStop}%
\bibitem [{\citenamefont {Ravensbergen}\ \emph
  {et~al.}(2018{\natexlab{a}})\citenamefont {Ravensbergen}, \citenamefont
  {Corre}, \citenamefont {Soave}, \citenamefont {Kreyer}, \citenamefont
  {Tzanova}, \citenamefont {Kiri\-lov},\ and\ \citenamefont
  {Grimm}}]{Ravensbergen2018ado}%
  \BibitemOpen
  \bibfield  {author} {\bibinfo {author} {\bibfnamefont {C.}~\bibnamefont
  {Ravensbergen}}, \bibinfo {author} {\bibfnamefont {V.}~\bibnamefont {Corre}},
  \bibinfo {author} {\bibfnamefont {E.}~\bibnamefont {Soave}}, \bibinfo
  {author} {\bibfnamefont {M.}~\bibnamefont {Kreyer}}, \bibinfo {author}
  {\bibfnamefont {S.}~\bibnamefont {Tzanova}}, \bibinfo {author} {\bibfnamefont
  {E.}~\bibnamefont {Kiri\-lov}}, \ and\ \bibinfo {author} {\bibfnamefont
  {R.}~\bibnamefont {Grimm}},\ }\href {\doibase 10.1103/PhysRevLett.120.223001}
  {\bibfield  {journal} {\bibinfo  {journal} {Phys. Rev. Lett.}\ }\textbf
  {\bibinfo {volume} {120}},\ \bibinfo {pages} {223001} (\bibinfo {year}
  {2018}{\natexlab{a}})}\BibitemShut {NoStop}%
\bibitem [{\citenamefont {Ravensbergen}\ \emph
  {et~al.}(2018{\natexlab{b}})\citenamefont {Ravensbergen}, \citenamefont
  {Corre}, \citenamefont {Soave}, \citenamefont {Kreyer}, \citenamefont
  {Kiri\-lov},\ and\ \citenamefont {Grimm}}]{Ravensbergen2018poa}%
  \BibitemOpen
  \bibfield  {author} {\bibinfo {author} {\bibfnamefont {C.}~\bibnamefont
  {Ravensbergen}}, \bibinfo {author} {\bibfnamefont {V.}~\bibnamefont {Corre}},
  \bibinfo {author} {\bibfnamefont {E.}~\bibnamefont {Soave}}, \bibinfo
  {author} {\bibfnamefont {M.}~\bibnamefont {Kreyer}}, \bibinfo {author}
  {\bibfnamefont {E.}~\bibnamefont {Kiri\-lov}}, \ and\ \bibinfo {author}
  {\bibfnamefont {R.}~\bibnamefont {Grimm}},\ }\href {\doibase
  10.1103/PhysRevA.98.063624} {\bibfield  {journal} {\bibinfo  {journal} {Phys.
  Rev. A}\ }\textbf {\bibinfo {volume} {98}},\ \bibinfo {pages} {063624}
  (\bibinfo {year} {2018}{\natexlab{b}})}\BibitemShut {NoStop}%
\bibitem [{\citenamefont {Petrov}\ \emph {et~al.}(2012)\citenamefont {Petrov},
  \citenamefont {Tiesinga},\ and\ \citenamefont {Kotochigova}}]{Petrov2012aif}%
  \BibitemOpen
  \bibfield  {author} {\bibinfo {author} {\bibfnamefont {A.}~\bibnamefont
  {Petrov}}, \bibinfo {author} {\bibfnamefont {E.}~\bibnamefont {Tiesinga}}, \
  and\ \bibinfo {author} {\bibfnamefont {S.}~\bibnamefont {Kotochigova}},\
  }\href {\doibase 10.1103/PhysRevLett.109.103002} {\bibfield  {journal}
  {\bibinfo  {journal} {Phys. Rev. Lett.}\ }\textbf {\bibinfo {volume} {109}},\
  \bibinfo {pages} {103002} (\bibinfo {year} {2012})}\BibitemShut {NoStop}%
\bibitem [{\citenamefont {Gonz\'alez-Mart\'{\i}nez}\ and\ \citenamefont
  {\ifmmode~\dot{Z}\else \.{Z}\fi{}uchowski}(2015)}]{Gonzalezmartinez2015mtf}%
  \BibitemOpen
  \bibfield  {author} {\bibinfo {author} {\bibfnamefont {M.~L.}\ \bibnamefont
  {Gonz\'alez-Mart\'{\i}nez}}\ and\ \bibinfo {author} {\bibfnamefont {P.~S.}\
  \bibnamefont {\ifmmode~\dot{Z}\else \.{Z}\fi{}uchowski}},\ }\href {\doibase
  10.1103/PhysRevA.92.022708} {\bibfield  {journal} {\bibinfo  {journal} {Phys.
  Rev. A}\ }\textbf {\bibinfo {volume} {92}},\ \bibinfo {pages} {022708}
  (\bibinfo {year} {2015})}\BibitemShut {NoStop}%
\bibitem [{SM()}]{SM}%
  \BibitemOpen
  \href@noop {} {}\bibinfo {note} {See Supplemental Material at xxxxxxxxxxxx
  for the experimental characterization of the Feshbach resonance scenario, the
  analysis of decay curves, and a model for the interaction-induced contraction
  of the mixture, which includes
  Refs.~\cite{Lange2009doa,Jachymski2013amo,OHara2002mot,Jochim2002mfc,Mosk2001mou,Taylorbook,Petrov2004tbp,Weber2003tbr,Lobo2006nso,Gezerlis2009hlf}.}\BibitemShut
  {Stop}%
\bibitem [{\citenamefont {Baumann}\ \emph {et~al.}(2014)\citenamefont
  {Baumann}, \citenamefont {Burdick}, \citenamefont {Lu},\ and\ \citenamefont
  {Lev}}]{Baumann2014ool}%
  \BibitemOpen
  \bibfield  {author} {\bibinfo {author} {\bibfnamefont {K.}~\bibnamefont
  {Baumann}}, \bibinfo {author} {\bibfnamefont {N.~Q.}\ \bibnamefont
  {Burdick}}, \bibinfo {author} {\bibfnamefont {M.}~\bibnamefont {Lu}}, \ and\
  \bibinfo {author} {\bibfnamefont {B.~L.}\ \bibnamefont {Lev}},\ }\href
  {\doibase 10.1103/PhysRevA.89.020701} {\bibfield  {journal} {\bibinfo
  {journal} {Phys. Rev. A}\ }\textbf {\bibinfo {volume} {89}},\ \bibinfo
  {pages} {020701} (\bibinfo {year} {2014})}\BibitemShut {NoStop}%
\bibitem [{\citenamefont {Burdick}\ \emph {et~al.}(2016)\citenamefont
  {Burdick}, \citenamefont {Tang},\ and\ \citenamefont {Lev}}]{Burdick2016lls}%
  \BibitemOpen
  \bibfield  {author} {\bibinfo {author} {\bibfnamefont {N.~Q.}\ \bibnamefont
  {Burdick}}, \bibinfo {author} {\bibfnamefont {Y.}~\bibnamefont {Tang}}, \
  and\ \bibinfo {author} {\bibfnamefont {B.~L.}\ \bibnamefont {Lev}},\ }\href
  {\doibase 10.1103/PhysRevX.6.031022} {\bibfield  {journal} {\bibinfo
  {journal} {Phys. Rev. X}\ }\textbf {\bibinfo {volume} {6}},\ \bibinfo {pages}
  {031022} (\bibinfo {year} {2016})}\BibitemShut {NoStop}%
\bibitem [{set()}]{settle}%
  \BibitemOpen
  \href@noop {} {}\bibinfo {note} {At 219.6\,G interspecies thermalization is
  sufficiently fast and Dy background losses show a pronounced
  minimum.}\BibitemShut {Stop}%
\bibitem [{the()}]{thermometry}%
  \BibitemOpen
  \href@noop {} {}\bibinfo {note} {Thermometry is based on time-of-flight
  images taken at high magnetic fields in regions where interspecies
  interactions are weak.}\BibitemShut {Stop}%
\bibitem [{odt()}]{odt}%
  \BibitemOpen
  \href@noop {} {}\bibinfo {note} {The ratio of the trap frequencies for K and
  Dy is essentially determined by the mass ratio and the polarizability ratio,
  which results in a $\bar{\omega}_{\rm K}/\bar{\omega}_{\rm Dy} = 3.60$
  \cite{Ravensbergen2018ado}.}\BibitemShut {Stop}%
\bibitem [{\citenamefont {Arndt}\ \emph {et~al.}(1997)\citenamefont {Arndt},
  \citenamefont {Dahan}, \citenamefont {Gu{\'e}ry-Odelin}, \citenamefont
  {Reynolds},\ and\ \citenamefont {Dalibard}}]{Arndt1997ooa}%
  \BibitemOpen
  \bibfield  {author} {\bibinfo {author} {\bibfnamefont {M.}~\bibnamefont
  {Arndt}}, \bibinfo {author} {\bibfnamefont {M.~B.}\ \bibnamefont {Dahan}},
  \bibinfo {author} {\bibfnamefont {D.}~\bibnamefont {Gu{\'e}ry-Odelin}},
  \bibinfo {author} {\bibfnamefont {M.}~\bibnamefont {Reynolds}}, \ and\
  \bibinfo {author} {\bibfnamefont {J.}~\bibnamefont {Dalibard}},\ }\href
  {\doibase 10.1103/PhysRevLett.79.625} {\bibfield  {journal} {\bibinfo
  {journal} {Phys. Rev. Lett.}\ }\textbf {\bibinfo {volume} {79}},\ \bibinfo
  {pages} {625} (\bibinfo {year} {1997})}\BibitemShut {NoStop}%
\bibitem [{\citenamefont {Gehm}\ \emph {et~al.}(2003)\citenamefont {Gehm},
  \citenamefont {Hemmer}, \citenamefont {O'Hara},\ and\ \citenamefont
  {Thomas}}]{Gehm2003ule}%
  \BibitemOpen
  \bibfield  {author} {\bibinfo {author} {\bibfnamefont {M.~E.}\ \bibnamefont
  {Gehm}}, \bibinfo {author} {\bibfnamefont {S.~L.}\ \bibnamefont {Hemmer}},
  \bibinfo {author} {\bibfnamefont {K.~M.}\ \bibnamefont {O'Hara}}, \ and\
  \bibinfo {author} {\bibfnamefont {J.~E.}\ \bibnamefont {Thomas}},\ }\href
  {\doibase 10.1103/PhysRevA.68.011603} {\bibfield  {journal} {\bibinfo
  {journal} {Phys. Rev. A}\ }\textbf {\bibinfo {volume} {68}},\ \bibinfo
  {pages} {011603} (\bibinfo {year} {2003})}\BibitemShut {NoStop}%
\bibitem [{col()}]{collrate}%
  \BibitemOpen
  \href@noop {} {}\bibinfo {note} {We estimate the collision rate for a K atom
  in the center of the Dy cloud by considering the resonant elastic scattering
  cross section $\sigma_{\rm res} = 4 \pi / \bar{k}^2_{\rm rel}$, the Dy peak
  number density $\hat{n}_{\rm Dy}$, and the mean relative velocity
  $\bar{v}_{\rm rel}$. For our typical conditions, $\hat{n}_{\rm Dy}
  \sigma_{\rm res} \bar{v}_{\rm rel} \approx 10^4\,$s$^{-1}$.}\BibitemShut
  {Stop}%
\bibitem [{\citenamefont {Anderlini}\ \emph {et~al.}(2005)\citenamefont
  {Anderlini}, \citenamefont {Ciampini}, \citenamefont {Cossart}, \citenamefont
  {Courtade}, \citenamefont {Cristiani}, \citenamefont {Sias}, \citenamefont
  {Morsch},\ and\ \citenamefont {Arimondo}}]{Anderlini2005mfc}%
  \BibitemOpen
  \bibfield  {author} {\bibinfo {author} {\bibfnamefont {M.}~\bibnamefont
  {Anderlini}}, \bibinfo {author} {\bibfnamefont {D.}~\bibnamefont {Ciampini}},
  \bibinfo {author} {\bibfnamefont {D.}~\bibnamefont {Cossart}}, \bibinfo
  {author} {\bibfnamefont {E.}~\bibnamefont {Courtade}}, \bibinfo {author}
  {\bibfnamefont {M.}~\bibnamefont {Cristiani}}, \bibinfo {author}
  {\bibfnamefont {C.}~\bibnamefont {Sias}}, \bibinfo {author} {\bibfnamefont
  {O.}~\bibnamefont {Morsch}}, \ and\ \bibinfo {author} {\bibfnamefont
  {E.}~\bibnamefont {Arimondo}},\ }\href {\doibase 10.1103/PhysRevA.72.033408}
  {\bibfield  {journal} {\bibinfo  {journal} {Phys. Rev. A}\ }\textbf {\bibinfo
  {volume} {72}},\ \bibinfo {pages} {033408} (\bibinfo {year}
  {2005})}\BibitemShut {NoStop}%
\bibitem [{\citenamefont {Regal}\ \emph {et~al.}(2004)\citenamefont {Regal},
  \citenamefont {Greiner},\ and\ \citenamefont {Jin}}]{Regal2004lom}%
  \BibitemOpen
  \bibfield  {author} {\bibinfo {author} {\bibfnamefont {C.~A.}\ \bibnamefont
  {Regal}}, \bibinfo {author} {\bibfnamefont {M.}~\bibnamefont {Greiner}}, \
  and\ \bibinfo {author} {\bibfnamefont {D.~S.}\ \bibnamefont {Jin}},\ }\href
  {\doibase 10.1103/PhysRevLett.92.083201} {\bibfield  {journal} {\bibinfo
  {journal} {Phys. Rev. Lett.}\ }\textbf {\bibinfo {volume} {92}},\ \bibinfo
  {eid} {083201} (\bibinfo {year} {2004})}\BibitemShut {NoStop}%
\bibitem [{Bfl()}]{Bfluct}%
  \BibitemOpen
  \href@noop {} {}\bibinfo {note} {Day-to-day fluctuations, drifts in the
  calibration, and residual ramping effects may cause magnetic-field
  uncertainties of the order of 100\,mG.}\BibitemShut {Stop}%
\bibitem [{\citenamefont {Dieckmann}\ \emph {et~al.}(2002)\citenamefont
  {Dieckmann}, \citenamefont {Stan}, \citenamefont {Gupta}, \citenamefont
  {Hadzibabic}, \citenamefont {Schunck},\ and\ \citenamefont
  {Ketterle}}]{Dieckmann2002doa}%
  \BibitemOpen
  \bibfield  {author} {\bibinfo {author} {\bibfnamefont {K.}~\bibnamefont
  {Dieckmann}}, \bibinfo {author} {\bibfnamefont {C.~A.}\ \bibnamefont {Stan}},
  \bibinfo {author} {\bibfnamefont {S.}~\bibnamefont {Gupta}}, \bibinfo
  {author} {\bibfnamefont {Z.}~\bibnamefont {Hadzibabic}}, \bibinfo {author}
  {\bibfnamefont {C.~H.}\ \bibnamefont {Schunck}}, \ and\ \bibinfo {author}
  {\bibfnamefont {W.}~\bibnamefont {Ketterle}},\ }\href {\doibase
  10.1103/PhysRevLett.89.203201} {\bibfield  {journal} {\bibinfo  {journal}
  {Phys. Rev. Lett.}\ }\textbf {\bibinfo {volume} {89}},\ \bibinfo {pages}
  {203201} (\bibinfo {year} {2002})}\BibitemShut {NoStop}%
\bibitem [{\citenamefont {Bourdel}\ \emph {et~al.}(2003)\citenamefont
  {Bourdel}, \citenamefont {Cubizolles}, \citenamefont {Khaykovich},
  \citenamefont {{Magalh\~aes}}, \citenamefont {Kokkelmans}, \citenamefont
  {Shlyapnikov},\ and\ \citenamefont {Salomon}}]{Bourdel2003mot}%
  \BibitemOpen
  \bibfield  {author} {\bibinfo {author} {\bibfnamefont {T.}~\bibnamefont
  {Bourdel}}, \bibinfo {author} {\bibfnamefont {J.}~\bibnamefont {Cubizolles}},
  \bibinfo {author} {\bibfnamefont {L.}~\bibnamefont {Khaykovich}}, \bibinfo
  {author} {\bibfnamefont {K.~M.~F.}\ \bibnamefont {{Magalh\~aes}}}, \bibinfo
  {author} {\bibfnamefont {S.~J. J. M.~F.}\ \bibnamefont {Kokkelmans}},
  \bibinfo {author} {\bibfnamefont {G.~V.}\ \bibnamefont {Shlyapnikov}}, \ and\
  \bibinfo {author} {\bibfnamefont {C.}~\bibnamefont {Salomon}},\ }\href
  {\doibase 10.1103/PhysRevLett.91.020402} {\bibfield  {journal} {\bibinfo
  {journal} {Phys. Rev. Lett.}\ }\textbf {\bibinfo {volume} {91}},\ \bibinfo
  {pages} {020402} (\bibinfo {year} {2003})}\BibitemShut {NoStop}%
\bibitem [{\citenamefont {Jochim}(2004)}]{Jochim2004PhD}%
  \BibitemOpen
  \bibfield  {author} {\bibinfo {author} {\bibfnamefont {S.}~\bibnamefont
  {Jochim}},\ }\emph {\bibinfo {title} {Bose-Einstein Condensation of
  Molecules}},\ \href@noop {} {Ph.D. thesis},\ \bibinfo  {school} {Innsbruck
  University} (\bibinfo {year} {2004})\BibitemShut {NoStop}%
\bibitem [{\citenamefont {Barontini}\ \emph {et~al.}(2009)\citenamefont
  {Barontini}, \citenamefont {Weber}, \citenamefont {Rabatti}, \citenamefont
  {Catani}, \citenamefont {Thalhammer}, \citenamefont {Inguscio},\ and\
  \citenamefont {Minardi}}]{Barontini2009ooh}%
  \BibitemOpen
  \bibfield  {author} {\bibinfo {author} {\bibfnamefont {G.}~\bibnamefont
  {Barontini}}, \bibinfo {author} {\bibfnamefont {C.}~\bibnamefont {Weber}},
  \bibinfo {author} {\bibfnamefont {F.}~\bibnamefont {Rabatti}}, \bibinfo
  {author} {\bibfnamefont {J.}~\bibnamefont {Catani}}, \bibinfo {author}
  {\bibfnamefont {G.}~\bibnamefont {Thalhammer}}, \bibinfo {author}
  {\bibfnamefont {M.}~\bibnamefont {Inguscio}}, \ and\ \bibinfo {author}
  {\bibfnamefont {F.}~\bibnamefont {Minardi}},\ }\href {\doibase
  10.1103/PhysRevLett.103.043201} {\bibfield  {journal} {\bibinfo  {journal}
  {Phys. Rev. Lett.}\ }\textbf {\bibinfo {volume} {103}},\ \bibinfo {pages}
  {043201} (\bibinfo {year} {2009})}\BibitemShut {NoStop}%
\bibitem [{\citenamefont {Maier}\ \emph {et~al.}(2015)\citenamefont {Maier},
  \citenamefont {Eisele}, \citenamefont {Tiemann},\ and\ \citenamefont
  {Zimmermann}}]{Maier2015era}%
  \BibitemOpen
  \bibfield  {author} {\bibinfo {author} {\bibfnamefont {R.~A.~W.}\
  \bibnamefont {Maier}}, \bibinfo {author} {\bibfnamefont {M.}~\bibnamefont
  {Eisele}}, \bibinfo {author} {\bibfnamefont {E.}~\bibnamefont {Tiemann}}, \
  and\ \bibinfo {author} {\bibfnamefont {C.}~\bibnamefont {Zimmermann}},\
  }\href {\doibase 10.1103/PhysRevLett.115.043201} {\bibfield  {journal}
  {\bibinfo  {journal} {Phys. Rev. Lett.}\ }\textbf {\bibinfo {volume} {115}},\
  \bibinfo {pages} {043201} (\bibinfo {year} {2015})}\BibitemShut {NoStop}%
\bibitem [{\citenamefont {Wacker}\ \emph {et~al.}(2016)\citenamefont {Wacker},
  \citenamefont {J\o{}rgensen}, \citenamefont {Birkmose}, \citenamefont
  {Winter}, \citenamefont {Mikkelsen}, \citenamefont {Sherson}, \citenamefont
  {Zinner},\ and\ \citenamefont {Arlt}}]{Wacker2016utb}%
  \BibitemOpen
  \bibfield  {author} {\bibinfo {author} {\bibfnamefont {L.~J.}\ \bibnamefont
  {Wacker}}, \bibinfo {author} {\bibfnamefont {N.~B.}\ \bibnamefont
  {J\o{}rgensen}}, \bibinfo {author} {\bibfnamefont {D.}~\bibnamefont
  {Birkmose}}, \bibinfo {author} {\bibfnamefont {N.}~\bibnamefont {Winter}},
  \bibinfo {author} {\bibfnamefont {M.}~\bibnamefont {Mikkelsen}}, \bibinfo
  {author} {\bibfnamefont {J.}~\bibnamefont {Sherson}}, \bibinfo {author}
  {\bibfnamefont {N.}~\bibnamefont {Zinner}}, \ and\ \bibinfo {author}
  {\bibfnamefont {J.~J.}\ \bibnamefont {Arlt}},\ }\href {\doibase
  10.1103/PhysRevLett.117.163201} {\bibfield  {journal} {\bibinfo  {journal}
  {Phys. Rev. Lett.}\ }\textbf {\bibinfo {volume} {117}},\ \bibinfo {pages}
  {163201} (\bibinfo {year} {2016})}\BibitemShut {NoStop}%
\bibitem [{\citenamefont {Bloom}\ \emph {et~al.}(2013)\citenamefont {Bloom},
  \citenamefont {Hu}, \citenamefont {Cumby},\ and\ \citenamefont
  {Jin}}]{Bloom2013tou}%
  \BibitemOpen
  \bibfield  {author} {\bibinfo {author} {\bibfnamefont {R.~S.}\ \bibnamefont
  {Bloom}}, \bibinfo {author} {\bibfnamefont {M.-G.}\ \bibnamefont {Hu}},
  \bibinfo {author} {\bibfnamefont {T.~D.}\ \bibnamefont {Cumby}}, \ and\
  \bibinfo {author} {\bibfnamefont {D.~S.}\ \bibnamefont {Jin}},\ }\href
  {\doibase 10.1103/PhysRevLett.111.105301} {\bibfield  {journal} {\bibinfo
  {journal} {Phys. Rev. Lett.}\ }\textbf {\bibinfo {volume} {111}},\ \bibinfo
  {pages} {105301} (\bibinfo {year} {2013})}\BibitemShut {NoStop}%
\bibitem [{\citenamefont {Pires}\ \emph {et~al.}(2014)\citenamefont {Pires},
  \citenamefont {Ulmanis}, \citenamefont {H\"afner}, \citenamefont {Repp},
  \citenamefont {Arias}, \citenamefont {Kuhnle},\ and\ \citenamefont
  {Weidem\"uller}}]{Pires2014ooe}%
  \BibitemOpen
  \bibfield  {author} {\bibinfo {author} {\bibfnamefont {R.}~\bibnamefont
  {Pires}}, \bibinfo {author} {\bibfnamefont {J.}~\bibnamefont {Ulmanis}},
  \bibinfo {author} {\bibfnamefont {S.}~\bibnamefont {H\"afner}}, \bibinfo
  {author} {\bibfnamefont {M.}~\bibnamefont {Repp}}, \bibinfo {author}
  {\bibfnamefont {A.}~\bibnamefont {Arias}}, \bibinfo {author} {\bibfnamefont
  {E.~D.}\ \bibnamefont {Kuhnle}}, \ and\ \bibinfo {author} {\bibfnamefont
  {M.}~\bibnamefont {Weidem\"uller}},\ }\href {\doibase
  10.1103/PhysRevLett.112.250404} {\bibfield  {journal} {\bibinfo  {journal}
  {Phys. Rev. Lett.}\ }\textbf {\bibinfo {volume} {112}},\ \bibinfo {pages}
  {250404} (\bibinfo {year} {2014})}\BibitemShut {NoStop}%
\bibitem [{\citenamefont {Tung}\ \emph {et~al.}(2014)\citenamefont {Tung},
  \citenamefont {Jim\'enez-Garc\'{\i}a}, \citenamefont {Johansen},
  \citenamefont {Parker},\ and\ \citenamefont {Chin}}]{Tung2014gso}%
  \BibitemOpen
  \bibfield  {author} {\bibinfo {author} {\bibfnamefont {S.-K.}\ \bibnamefont
  {Tung}}, \bibinfo {author} {\bibfnamefont {K.}~\bibnamefont
  {Jim\'enez-Garc\'{\i}a}}, \bibinfo {author} {\bibfnamefont {J.}~\bibnamefont
  {Johansen}}, \bibinfo {author} {\bibfnamefont {C.~V.}\ \bibnamefont
  {Parker}}, \ and\ \bibinfo {author} {\bibfnamefont {C.}~\bibnamefont
  {Chin}},\ }\href {\doibase 10.1103/PhysRevLett.113.240402} {\bibfield
  {journal} {\bibinfo  {journal} {Phys. Rev. Lett.}\ }\textbf {\bibinfo
  {volume} {113}},\ \bibinfo {pages} {240402} (\bibinfo {year}
  {2014})}\BibitemShut {NoStop}%
\bibitem [{\citenamefont {Ulmanis}\ \emph {et~al.}(2016)\citenamefont
  {Ulmanis}, \citenamefont {H\"afner}, \citenamefont {Pires}, \citenamefont
  {Werner}, \citenamefont {Petrov}, \citenamefont {Kuhnle},\ and\ \citenamefont
  {Weidem\"uller}}]{Ulmanis2016utb}%
  \BibitemOpen
  \bibfield  {author} {\bibinfo {author} {\bibfnamefont {J.}~\bibnamefont
  {Ulmanis}}, \bibinfo {author} {\bibfnamefont {S.}~\bibnamefont {H\"afner}},
  \bibinfo {author} {\bibfnamefont {R.}~\bibnamefont {Pires}}, \bibinfo
  {author} {\bibfnamefont {F.}~\bibnamefont {Werner}}, \bibinfo {author}
  {\bibfnamefont {D.~S.}\ \bibnamefont {Petrov}}, \bibinfo {author}
  {\bibfnamefont {E.~D.}\ \bibnamefont {Kuhnle}}, \ and\ \bibinfo {author}
  {\bibfnamefont {M.}~\bibnamefont {Weidem\"uller}},\ }\href {\doibase
  10.1103/PhysRevA.93.022707} {\bibfield  {journal} {\bibinfo  {journal} {Phys.
  Rev. A}\ }\textbf {\bibinfo {volume} {93}},\ \bibinfo {pages} {022707}
  (\bibinfo {year} {2016})}\BibitemShut {NoStop}%
\bibitem [{\citenamefont {Lous}\ \emph {et~al.}(2018)\citenamefont {Lous},
  \citenamefont {Fritsche}, \citenamefont {Jag}, \citenamefont {Lehmann},
  \citenamefont {Kirilov}, \citenamefont {Huang},\ and\ \citenamefont
  {Grimm}}]{Lous2018pti}%
  \BibitemOpen
  \bibfield  {author} {\bibinfo {author} {\bibfnamefont {R.~S.}\ \bibnamefont
  {Lous}}, \bibinfo {author} {\bibfnamefont {I.}~\bibnamefont {Fritsche}},
  \bibinfo {author} {\bibfnamefont {M.}~\bibnamefont {Jag}}, \bibinfo {author}
  {\bibfnamefont {F.}~\bibnamefont {Lehmann}}, \bibinfo {author} {\bibfnamefont
  {E.}~\bibnamefont {Kirilov}}, \bibinfo {author} {\bibfnamefont
  {B.}~\bibnamefont {Huang}}, \ and\ \bibinfo {author} {\bibfnamefont
  {R.}~\bibnamefont {Grimm}},\ }\href {\doibase 10.1103/PhysRevLett.120.243403}
  {\bibfield  {journal} {\bibinfo  {journal} {Phys. Rev. Lett.}\ }\textbf
  {\bibinfo {volume} {120}},\ \bibinfo {pages} {243403} (\bibinfo {year}
  {2018})}\BibitemShut {NoStop}%
\bibitem [{\citenamefont {Tzanova}(tion)}]{TzanovaPhD}%
  \BibitemOpen
  \bibfield  {author} {\bibinfo {author} {\bibfnamefont {S.}~\bibnamefont
  {Tzanova}},\ }\href@noop {} {Ph.D. thesis},\ \bibinfo  {school} {University
  of Innsbruck} (\bibinfo {year} {in preparation})\BibitemShut {NoStop}%
\bibitem [{\citenamefont {Zwierlein}\ \emph {et~al.}(2006)\citenamefont
  {Zwierlein}, \citenamefont {Schirotzek}, \citenamefont {Schunck},\ and\
  \citenamefont {Ketterle}}]{Zwierlein2006doo}%
  \BibitemOpen
  \bibfield  {author} {\bibinfo {author} {\bibfnamefont {M.~W.}\ \bibnamefont
  {Zwierlein}}, \bibinfo {author} {\bibfnamefont {A.}~\bibnamefont
  {Schirotzek}}, \bibinfo {author} {\bibfnamefont {C.~H.}\ \bibnamefont
  {Schunck}}, \ and\ \bibinfo {author} {\bibfnamefont {W.}~\bibnamefont
  {Ketterle}},\ }\href {\doibase doi:10.1038/nature04936} {\bibfield  {journal}
  {\bibinfo  {journal} {Nature (London)}\ }\textbf {\bibinfo {volume} {442}},\
  \bibinfo {pages} {54} (\bibinfo {year} {2006})}\BibitemShut {NoStop}%
\bibitem [{\citenamefont {Baarsma}\ \emph {et~al.}(2010)\citenamefont
  {Baarsma}, \citenamefont {Gubbels},\ and\ \citenamefont
  {Stoof}}]{Baarsma2010pam}%
  \BibitemOpen
  \bibfield  {author} {\bibinfo {author} {\bibfnamefont {J.~E.}\ \bibnamefont
  {Baarsma}}, \bibinfo {author} {\bibfnamefont {K.~B.}\ \bibnamefont
  {Gubbels}}, \ and\ \bibinfo {author} {\bibfnamefont {H.~T.~C.}\ \bibnamefont
  {Stoof}},\ }\href {\doibase 10.1103/PhysRevA.82.013624} {\bibfield  {journal}
  {\bibinfo  {journal} {Phys. Rev. A}\ }\textbf {\bibinfo {volume} {82}},\
  \bibinfo {pages} {013624} (\bibinfo {year} {2010})}\BibitemShut {NoStop}%
\bibitem [{lif()}]{lifshitz}%
  \BibitemOpen
  \href@noop {} {}\bibinfo {note} {Note that in Ref.~\cite{Gubbels2009lpi}
  temperatures are given in units of a reduced Fermi temperature, which at the
  Lifshitz point is a factor of 3.4 higher than the Fermi temperature of the
  heavy species.}\BibitemShut {Stop}%
\bibitem [{\citenamefont {Lange}\ \emph {et~al.}(2009)\citenamefont {Lange},
  \citenamefont {Pilch}, \citenamefont {Prantner}, \citenamefont {Ferlaino},
  \citenamefont {Engeser}, \citenamefont {N\"{a}gerl}, \citenamefont {Grimm},\
  and\ \citenamefont {Chin}}]{Lange2009doa}%
  \BibitemOpen
  \bibfield  {author} {\bibinfo {author} {\bibfnamefont {A.~D.}\ \bibnamefont
  {Lange}}, \bibinfo {author} {\bibfnamefont {K.}~\bibnamefont {Pilch}},
  \bibinfo {author} {\bibfnamefont {A.}~\bibnamefont {Prantner}}, \bibinfo
  {author} {\bibfnamefont {F.}~\bibnamefont {Ferlaino}}, \bibinfo {author}
  {\bibfnamefont {B.}~\bibnamefont {Engeser}}, \bibinfo {author} {\bibfnamefont
  {H.-C.}\ \bibnamefont {N\"{a}gerl}}, \bibinfo {author} {\bibfnamefont
  {R.}~\bibnamefont {Grimm}}, \ and\ \bibinfo {author} {\bibfnamefont
  {C.}~\bibnamefont {Chin}},\ }\href {\doibase 10.1103/PhysRevA.79.013622}
  {\bibfield  {journal} {\bibinfo  {journal} {Phys. Rev. A}\ }\textbf {\bibinfo
  {volume} {79}},\ \bibinfo {pages} {013622} (\bibinfo {year}
  {2009})}\BibitemShut {NoStop}%
\bibitem [{\citenamefont {Jachymski}\ and\ \citenamefont
  {Julienne}(2013)}]{Jachymski2013amo}%
  \BibitemOpen
  \bibfield  {author} {\bibinfo {author} {\bibfnamefont {K.}~\bibnamefont
  {Jachymski}}\ and\ \bibinfo {author} {\bibfnamefont {P.~S.}\ \bibnamefont
  {Julienne}},\ }\href {\doibase 10.1103/PhysRevA.88.052701} {\bibfield
  {journal} {\bibinfo  {journal} {Phys. Rev. A}\ }\textbf {\bibinfo {volume}
  {88}},\ \bibinfo {pages} {052701} (\bibinfo {year} {2013})}\BibitemShut
  {NoStop}%
\bibitem [{\citenamefont {O'Hara}\ \emph {et~al.}(2002)\citenamefont {O'Hara},
  \citenamefont {Hemmer}, \citenamefont {Granade}, \citenamefont {Gehm},
  \citenamefont {Thomas}, \citenamefont {Venturi}, \citenamefont {Tiesinga},\
  and\ \citenamefont {Williams}}]{OHara2002mot}%
  \BibitemOpen
  \bibfield  {author} {\bibinfo {author} {\bibfnamefont {K.~M.}\ \bibnamefont
  {O'Hara}}, \bibinfo {author} {\bibfnamefont {S.~L.}\ \bibnamefont {Hemmer}},
  \bibinfo {author} {\bibfnamefont {S.~R.}\ \bibnamefont {Granade}}, \bibinfo
  {author} {\bibfnamefont {M.~E.}\ \bibnamefont {Gehm}}, \bibinfo {author}
  {\bibfnamefont {J.~E.}\ \bibnamefont {Thomas}}, \bibinfo {author}
  {\bibfnamefont {V.}~\bibnamefont {Venturi}}, \bibinfo {author} {\bibfnamefont
  {E.}~\bibnamefont {Tiesinga}}, \ and\ \bibinfo {author} {\bibfnamefont
  {C.~J.}\ \bibnamefont {Williams}},\ }\href {\doibase
  10.1103/PhysRevA.66.041401} {\bibfield  {journal} {\bibinfo  {journal} {Phys.
  Rev. A}\ }\textbf {\bibinfo {volume} {66}},\ \bibinfo {pages} {041401}
  (\bibinfo {year} {2002})}\BibitemShut {NoStop}%
\bibitem [{\citenamefont {Jochim}\ \emph {et~al.}(2002)\citenamefont {Jochim},
  \citenamefont {Bartenstein}, \citenamefont {Hendl}, \citenamefont {{Hecker
  Denschlag}}, \citenamefont {Grimm}, \citenamefont {Mosk},\ and\ \citenamefont
  {Weidem\"uller}}]{Jochim2002mfc}%
  \BibitemOpen
  \bibfield  {author} {\bibinfo {author} {\bibfnamefont {S.}~\bibnamefont
  {Jochim}}, \bibinfo {author} {\bibfnamefont {M.}~\bibnamefont {Bartenstein}},
  \bibinfo {author} {\bibfnamefont {G.}~\bibnamefont {Hendl}}, \bibinfo
  {author} {\bibfnamefont {J.}~\bibnamefont {{Hecker Denschlag}}}, \bibinfo
  {author} {\bibfnamefont {R.}~\bibnamefont {Grimm}}, \bibinfo {author}
  {\bibfnamefont {A.}~\bibnamefont {Mosk}}, \ and\ \bibinfo {author}
  {\bibfnamefont {W.}~\bibnamefont {Weidem\"uller}},\ }\href {\doibase
  10.1103/PhysRevLett.89.273202} {\bibfield  {journal} {\bibinfo  {journal}
  {Phys. Rev. Lett}\ }\textbf {\bibinfo {volume} {89}},\ \bibinfo {pages}
  {273202} (\bibinfo {year} {2002})}\BibitemShut {NoStop}%
\bibitem [{\citenamefont {Mosk}\ \emph {et~al.}(2001)\citenamefont {Mosk},
  \citenamefont {Kraft}, \citenamefont {Mudrich}, \citenamefont {Singer},
  \citenamefont {Wohlleben}, \citenamefont {Grimm},\ and\ \citenamefont
  {Weidem\"uller}}]{Mosk2001mou}%
  \BibitemOpen
  \bibfield  {author} {\bibinfo {author} {\bibfnamefont {A.}~\bibnamefont
  {Mosk}}, \bibinfo {author} {\bibfnamefont {S.}~\bibnamefont {Kraft}},
  \bibinfo {author} {\bibfnamefont {M.}~\bibnamefont {Mudrich}}, \bibinfo
  {author} {\bibfnamefont {K.}~\bibnamefont {Singer}}, \bibinfo {author}
  {\bibfnamefont {W.}~\bibnamefont {Wohlleben}}, \bibinfo {author}
  {\bibfnamefont {R.}~\bibnamefont {Grimm}}, \ and\ \bibinfo {author}
  {\bibfnamefont {M.}~\bibnamefont {Weidem\"uller}},\ }\href {\doibase
  10.1007/s003400100743} {\bibfield  {journal} {\bibinfo  {journal} {Appl.
  Phys. B}\ }\textbf {\bibinfo {volume} {73}},\ \bibinfo {pages} {791}
  (\bibinfo {year} {2001})}\BibitemShut {NoStop}%
\bibitem [{\citenamefont {Taylor}(1997)}]{Taylorbook}%
  \BibitemOpen
  \bibfield  {author} {\bibinfo {author} {\bibfnamefont {J.~R.}\ \bibnamefont
  {Taylor}},\ }\href@noop {} {\emph {\bibinfo {title} {An Introduction to Error
  Analysis}}}\ (\bibinfo  {publisher} {University Science Books},\ \bibinfo
  {year} {1997})\BibitemShut {NoStop}%
\bibitem [{\citenamefont {Petrov}(2004)}]{Petrov2004tbp}%
  \BibitemOpen
  \bibfield  {author} {\bibinfo {author} {\bibfnamefont {D.~S.}\ \bibnamefont
  {Petrov}},\ }\href {\doibase 10.1103/PhysRevLett.93.143201} {\bibfield
  {journal} {\bibinfo  {journal} {Phys. Rev. Lett.}\ }\textbf {\bibinfo
  {volume} {93}},\ \bibinfo {pages} {143201} (\bibinfo {year}
  {2004})}\BibitemShut {NoStop}%
\bibitem [{\citenamefont {Weber}\ \emph {et~al.}(2003)\citenamefont {Weber},
  \citenamefont {Herbig}, \citenamefont {Mark}, \citenamefont {N\"agerl},\ and\
  \citenamefont {Grimm}}]{Weber2003tbr}%
  \BibitemOpen
  \bibfield  {author} {\bibinfo {author} {\bibfnamefont {T.}~\bibnamefont
  {Weber}}, \bibinfo {author} {\bibfnamefont {J.}~\bibnamefont {Herbig}},
  \bibinfo {author} {\bibfnamefont {M.}~\bibnamefont {Mark}}, \bibinfo {author}
  {\bibfnamefont {H.-C.}\ \bibnamefont {N\"agerl}}, \ and\ \bibinfo {author}
  {\bibfnamefont {R.}~\bibnamefont {Grimm}},\ }\href {\doibase
  10.1103/PhysRevLett.91.123201} {\bibfield  {journal} {\bibinfo  {journal}
  {Phys. Rev. Lett.}\ }\textbf {\bibinfo {volume} {91}},\ \bibinfo {pages}
  {123201} (\bibinfo {year} {2003})}\BibitemShut {NoStop}%
\bibitem [{\citenamefont {Lobo}\ \emph {et~al.}(2006)\citenamefont {Lobo},
  \citenamefont {Recati}, \citenamefont {Giorgini},\ and\ \citenamefont
  {Stringari}}]{Lobo2006nso}%
  \BibitemOpen
  \bibfield  {author} {\bibinfo {author} {\bibfnamefont {C.}~\bibnamefont
  {Lobo}}, \bibinfo {author} {\bibfnamefont {A.}~\bibnamefont {Recati}},
  \bibinfo {author} {\bibfnamefont {S.}~\bibnamefont {Giorgini}}, \ and\
  \bibinfo {author} {\bibfnamefont {S.}~\bibnamefont {Stringari}},\ }\href
  {\doibase 10.1103/PhysRevLett.97.200403} {\bibfield  {journal} {\bibinfo
  {journal} {Phys. Rev. Lett.}\ }\textbf {\bibinfo {volume} {97}},\ \bibinfo
  {pages} {200403} (\bibinfo {year} {2006})}\BibitemShut {NoStop}%
\bibitem [{\citenamefont {Gezerlis}\ \emph {et~al.}(2009)\citenamefont
  {Gezerlis}, \citenamefont {Gandolfi}, \citenamefont {Schmidt},\ and\
  \citenamefont {Carlson}}]{Gezerlis2009hlf}%
  \BibitemOpen
  \bibfield  {author} {\bibinfo {author} {\bibfnamefont {A.}~\bibnamefont
  {Gezerlis}}, \bibinfo {author} {\bibfnamefont {S.}~\bibnamefont {Gandolfi}},
  \bibinfo {author} {\bibfnamefont {K.~E.}\ \bibnamefont {Schmidt}}, \ and\
  \bibinfo {author} {\bibfnamefont {J.}~\bibnamefont {Carlson}},\ }\href
  {\doibase 10.1103/PhysRevLett.103.060403} {\bibfield  {journal} {\bibinfo
  {journal} {Phys. Rev. Lett.}\ }\textbf {\bibinfo {volume} {103}},\ \bibinfo
  {pages} {060403} (\bibinfo {year} {2009})}\BibitemShut {NoStop}%
\end{thebibliography}

%

\end{document}